\documentclass[11pt]{article}
\usepackage{graphicx}
\usepackage{amsmath}
\newcommand\lra{\leftrightarrow}
\newcommand\tr{{\rm tr}}
\newcommand{\Ref}[1]{(\ref{#1})}

\newcommand\beq{\begin{equation}}
\newcommand\eeq{\end{equation}}
\newcommand\bl{{\bf l}}
\newcommand\bb{{\bf b}}

\title{\bf Gluon production in the Lipatov effective action formalism.}
\author{M.A.Braun, S.S.Pozdnyakov, M.Yu.Salykin, M.I.Vyazovsky\\
Dep. of High Energy physics,
 Saint-Petersburg State University,\\
198504 S.Petersburg, Russia}

\begin{document}

\maketitle

\noindent {\Large\bf Abstract}

Gluon production on two scattering centers is studied in the
formalism of reggeized gluons. Different contributions to the
inclusive cross-section are derived with the help of the Lipatov
effective action. The AGK relations between these contributions are
established. The found inclusive cross-section is compared to the
one in the dipole picture and demonstrated to be the same.

%%%%%%%%%%%%%%%%%%%%%%%%%%%%%%%%%%%%%%%%%%
%%%%%%%%%%%%%%%%%%%%%%%%%%%%%%%%%%%%%%%%%%
%%%%%%%%%%%%%%%%%%%%%%%%%%%%%%%%%%%%%%%%%%

\section{Introduction}
One of the
basic processes in high-energy collisions is the
inclusive gluon production off  heavy nuclear targets.
At high energies, in the QCD with a large number of colours $N_c\to\infty$
it can be studied either in the
interacting reggeized gluon approach ~\cite{braun1,braun3,BSVC}
or, alternatively, in the framework
of the dipole picture (or equivalent  JIMWLK approach with certain
approximations) in which the interacting hadrons are presented in terms
of  colour dipoles with their density evolving in rapidity ~\cite{KT}.
The two approaches seem to be based on different pictures and approximations
and it is vitally important to understand if they are completely equivalent
or have some significant differences. It is well-known that the total
cross-section for the dipole-dipole scattering turns out to be identical
in the reggeized gluons and  evolving dipoles pictures, both approaches
leading to the same BFKL equation, although in different spaces,
momentum or coordinate ones.
The situation with the inclusive gluon production turned out to be more
complicated. In particular in ~\cite{BSVC} it was advocated that
this cross-section off two centers found in the
framework of the dipole approach in ~\cite{KT} is not complete and
has to be supplemented by new terms involving states composed of three
and four reggeized gluons (the so-called BKP states ~\cite{bartels,kwie}).
On the other
hand in ~\cite{braun2} it was shown that, at least in
lowest orders, contributions from such states in fact cancel, so that
one is left with
exactly the cross-section obtained in the dipole approach.
It should be stressed however that this conclusion was found in the
purely transversal approach in which it the validity
of the standard AGK relations
~\cite{AGK} for different cuts of the scattering amplitude was assumed
without proof.
To finally compare  the
reggeized gluon and dipole approaches one has to study contributions
from all different cuts and check the validity of the mentioned AGK
rules. This cannot be done in the purely transverse approach used
in ~\cite{braun1,braun3,braun2} (nor in the dipole approach) but requires
knowledge of the amplitude
as a function of its longitudinal momenta. This information is trivial
for production  on  a single scattering center in the target. But it
becomes considerably more complicated when the target involves two or more
centers.

To study the inclusive gluon production taking into account the dependence
of the relevant amplitudes on the longitudinal momenta we use
the Lipatov effective action ~\cite{lipatov}, which provides
a powerful and constructive technique for the calculation of
all Feynman diagrams in the Regge kinematics.

\begin{figure}[h]
\begin{center}
\includegraphics[scale=0.60]{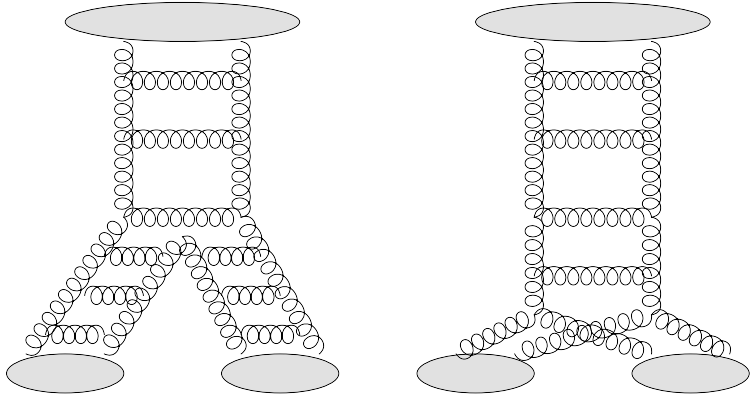}
\end{center}
\caption{Diagrams for the total cross-section on two centers.}
\label{npap3}
\end{figure}

To understand our derivation it is instructive to first visualize the
elastic amplitude for the scattering off the nuclear target in the
reggeized gluon technique, which, as mentioned is completely equivalent to
the dipole picture. The bulk of the amplitude can then be presented in
terms of pomerons propagating from the projectile to many nucleons of the
target each one splitting in two with a certain known triple pomeron vertex
(pomeron fan diagrams). These diagrams are to be supplemented by simpler
ones with less pomerons, which contribution corresponds to the
glauberized initial condition for the dipole-nucleus amplitude in the
dipole formalism. These two parts are shown in Fig. \ref{npap3}
for the scattering on two centers.

\begin{figure}[h]
\begin{center}
\includegraphics[scale=0.67]{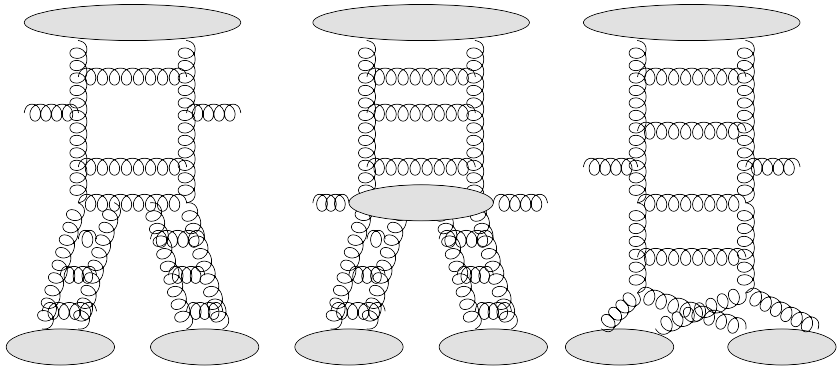}
\end{center}
\caption{Diagrams for the inclusive cross-section on two centers.}
\label{npap4}
\end{figure}

To calculate the inclusive cross-section one has to fix a real intermediate
gluon either inside the initial pomeron or inside the triple pomeron vertex.
The contribution from the real gluons inside the pomerons after splitting
is absent due to trivial AGK cancellations. So the inclusive cross-section
actually consists of three parts: emission from the initial pomeron before all
splittings, emission from the triple pomeron vertex (Fig. \ref{npap3}, the
first diagram) and emission from the pomeron directly coupled to the targets
(Fig. \ref{npap3}, the second diagram).
These three parts are illustrated in Fig. \ref{npap4}.
Emission from the pomeron chain is well understood and has the same form
in the reggeized gluon picture and the dipole model.
Thus the comparison of both approaches is reduced to
calculation of the contribution from the triple pomeron vertex. Since the
latter does not involve evolution and so contains only one intermediate gluon
this comparison can be made in the lowest order in the coupling constant
and moreover for any choice of the projectile and targets.
This allows to simplify the problem and reduce it to calculation of the
inclusive cross-section on only two centers in the lowest order with the
projectile and two targets chosen as quarks. This
calculation constitutes the subject of our paper.

\begin{figure}[h]
\begin{center}
\includegraphics[scale=0.40]{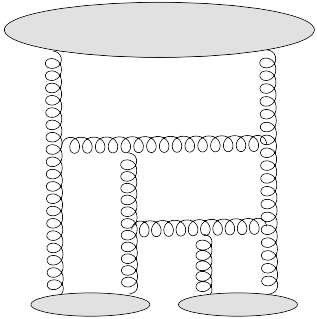}
\end{center}
\caption{Diagram with consecutive splitting of the initial pomeron.}
\label{fig2a}
\end{figure}

Note that one can additionally draw diagrams with consecutive splitting
of the initial pomeron into first three and afterwards four reggeized
gluons, as shown in Fig. \ref{fig2a}.
As was shown long ago, their contribution in fact reduces to that
of the diagrams in Fig. \ref{npap3} with the effective 3-pomeron vertex
~\cite{BW}. However, there remained a question whether such a reduction
worked for the inclusive cross-sections. Precisely this problem was
discussed in ~\cite{BSVC}. As mentioned above in ~\cite{braun2} in the
lowest non-trivial (next-to-leading) order it was shown that diagrams
of the type depicted in Fig. \ref{fig2a} give no new contributions.
In the dipole approach this result is valid at all orders ~\cite{KT}.
For this reason we do not consider such diagrams.

As mentioned the important tool in our calculation is the
Lipatov effective action which allows to calculate the vertices for
transition of a reggeized gluon (''reggeon'' (R)) into one, two or three
reggeons with emission of a real gluon (''particle'' (P)), that is
transitions R$\to$RP, R$\to$RRP, R$\to$RRRP. While the R$\to$RP (''Lipatov'')
vertex has been known since long ago, the  required R$\to$RRP and
R$\to$RRRP vertices were recently calculated in
papers ~\cite{BV} and ~\cite{BPSV} respectively. A remarkable result derived in
~\cite{BLSV} for the  R$\to$RRP vertex, and
in ~\cite{BPSV} for the R$\to$RRRP is that the 4 dimensional
amplitudes calculated with the help
of the effective action can be obtained from the purely transverse picture
with real intermediate particles (quarks and gluons) carrying the
standard Feynman propagators (see ~\cite{BSV} for  details).
This technique will be the main instrument
in our calculations of the inclusive cross-section.

\begin{figure}[h]
\begin{center}
\includegraphics[scale=0.60]{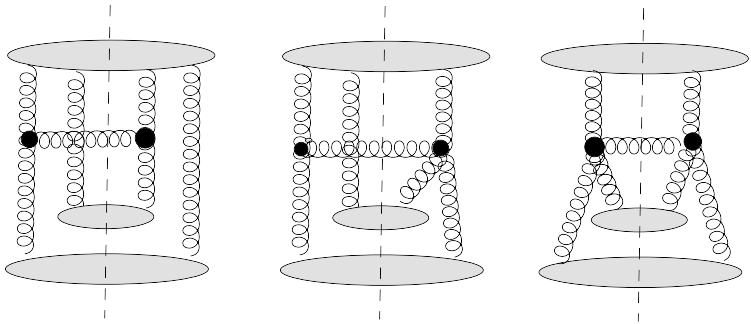}
\end{center}
\caption{Typical diagrams for the inclusive cross-section on two centers
with both target quarks in the intermediate states.}
\label{npap5}
\end{figure}

We have also to mention that a part of the inclusive cross-section
corresponding to the intermediate states with real quarks from both of
the targets (see Fig. \ref{npap5}) has already been calculated in
~\cite{BSV}. So in this paper we calculate the rest of the contributions:
the diffractive one (D) and that with only one of the target quarks
appearing as real in the intermediate states (''single cut'' (SC)). After
summing all the contributions we compare the result with the dipole picture.
Our conclusion is that the inclusive cross-section for gluon production
is the same in both approaches.

The paper is organized as follows.
Sections 2 and 3 are devoted to the calculations of SC and D contributions,
respectively. In Section 4 we sum all the contribution to find the final
expression for the inclusive cross-section and compare our result
with the one in the dipole model in the same order to establish their identity.
Finally we present our conclusions in the last section.
Our appendices serve to establish correspondence of
our diagrams with the cross-section and recall the derivation
of the inclusive cross-section for gluon production from the BFKL chain.

\section{''Single cut'' contribution}
\subsection{General notation}

\begin{figure}[h]
\begin{center}
\includegraphics[scale=0.55]{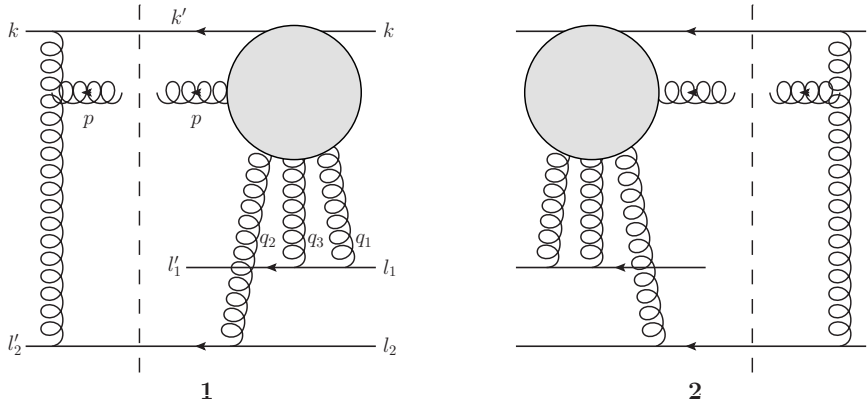}
\end{center}
\caption{Diagrams of the ''single cut'' contribution.}
\label{FC}
\end{figure}

We consider the imaginary part of the amplitude for production
of a gluon $p$ in the collision
of the projectile quark with equal initial and final momenta $k$
with two target quarks with  initial momenta $l_1=l_2=l$
and  final momenta $l'_1,l'_2$,\ \  $l'_1+l'_2=l_1+l_2$
(see Fig.~\ref{FC},1).
We denote $k'=k-q$ the intermediate quark momentum.
The momentum transferred to the targets is $\lambda=l_1'-l_1=l_2-l_2'$.
To the total result one has to add
the complex conjugated contribution corresponding to Fig.~\ref{FC},2
and a similar  amplitude with the exchanged targets $l_1$ and $l_2$.
The latter gives the same contribution and so can be dropped
for the imaginary part which is one half of the discontinuity.

The general Regge kinematics with the emitted gluon momentum $p$
in the central region is defined by the following relations:
$$
k_+>>p_+\approx q_+>>q_{1,2,3+}\sim l_{1,2+}\sim l'_{1,2+} \ ,
$$
\begin{equation}
l_{1,2-}\sim l'_{1,2-}>>q_{1,2,3-}\sim p_- >> q_{-}\sim k_- \ ,
\label{e1}
\end{equation}
whereas all transverse components are implied to be of the same order
much smaller than the largest longitudinal components.
As shown in Appendix 1, the leading contribution to the production
on two targets from the heavy nucleus is determined
only by the delta-functional singularity at
$2(k\lambda)=\sqrt{s}\lambda_{-}=0$ with
$
\lambda_+=0,\ \lambda_{\perp}=0.
$
We work in the c.m. system where
\begin{equation}
k_+=\sqrt{s},\quad k_-=0,\quad l_-=\sqrt{s},\quad l_+=0,
\quad k_{\perp}=l_{\perp}=0;
\quad s=(k+l)^2 \ .
\end{equation}

In the Regge approximation one have to neglect all transverse components
of momenta compared to large longitudinal ones.
The relative order of components of different momenta was considered
in \cite{BSV} and it was found that one has to neglect also
the component $p_-$ (which equals $-p^2_{\perp}/p_+$ for real gluon)
compared to $\lambda_-$ before separating out
the terms singular in the point $\lambda_- =0$.
These rules are equivalent to taking the limit
$k_+,p_+ \to\infty$, $k_+ >> p_+$.

The gluon production amplitude with three reggeons attached to the targets
(the filled blob on the right side of the cut in Fig.~\ref{FC},1)
was found in \cite{BPSV} as a sum of three terms:
$$
-g^{4}\gamma_{+}f^{db_3 a}
\left[ \frac{k_{+}^{2}\cdot T^{d}T^{b_2}T^{b_1}}
 { ((k-q_{1}-q_{2})^{2}+i0) ((k-q_{1})^{2}+i0) }
+ \frac{k_{+}^{2}\cdot T^{b_1}T^{d}T^{b_2}}
 { ((k'+q_{1})^{2}+i0) ((k-q_{2})^{2}+i0)}
\right.
$$
\begin{equation}
+ \left. \frac{k_{+}^{2}\cdot T^{b_{2}}T^{b_{1}}T^{d}}
 { ((k'+q_{2})^{2}+i0) ((k'+q_{1}+q_{2})^{2}+i0) } \right]
L(p,q_3)+{\cal P}
\label{ea3}
\end{equation}
and
\begin{equation}
g^{4}  \gamma_{+} f^{db_{3}a}
\frac{q_{+}B(p,q_{3},q_{2})}{(q-q_{1}-q_{2})^{2}+i0}\!
\left[ \frac{k_{+}\cdot if^{cb_2 d}T^{c}T^{b_1}}{(k-q_{1})^{2}+i0}
+\frac{k_{+}\cdot if^{cb_2 d}T^{b_1}T^{c}}{(k'+q_{1})^{2}+i0}
\right]+{\cal P}
\label{ea2}
\end{equation}
and
\begin{equation}
g^{4} \gamma_{+} T^{b}f^{bb_{1}c}f^{cb_{2}d}f^{db_{3}a}
\frac{q_{+}^{2}\, B(p,q_{3}+q_{2},q_{1})}
{((q-q_{1})^{2}+i0)((q-q_{1}-q_{2})^{2}+i0)}+{\cal P}
\label{ea1}
\end{equation}
with the common factor
$8/q^2_{1\perp}q^2_{2\perp}q^2_{3\perp}$.
Here ${\cal P}$ means permutations of $ q_1,q_2,q_3$ and $b_1,b_2,b_3$.
\begin{equation}
L(p,q)=\frac{(p\epsilon_{\perp})}{p_\perp^2}-
\frac{(p+q,\epsilon_{\perp})}{(p+q)_\perp^2}
\label{evl}
\end{equation}
and
\begin{equation}
B(p,q_{2},q_{1})=L(p+q_2,q_1)
=\frac{(p+q_2,\epsilon_{\perp})}{(p+q_2)_\perp^2}
- \frac{(p+q_2+q_1,\epsilon_{\perp})}{(p+q_2+q_1)_\perp^2}
\label{evb}
\end{equation}
are, respectively, the Lipatov vertex which describes gluon production
from a reggeon and the Bartels vertex which describes gluon production
from a reggeon splitting to two or three reggeons.

The conjugated amplitude on the left side of the cut in Fig.~\ref{FC},1
is
\begin{equation}
-i g^3 [\bar{u}({\bf k}) \gamma_{+} u({\bf k'})]\cdot
[\bar{u}({\bf l'_2}) \gamma_{-} u({\bf l''_2})]
\frac{f^{b' b_4 a} T^{b'} L(p,q_2)}{q_{2\perp}^2} \ .
\label{e11}
\end{equation}

Diagrams for ''single cut'' contribution are shown
in Figs.~\ref{C1}-\ref{A1}.
The notation of momenta is as follows.
The target interacting with two reggeons (uncut line)
has the momentum $l_1$, the correspondent reggeon momentum flowing
into the right interaction vertex is denoted $q_1$ for all cases
and $q_3$ is the momentum flowing into the left vertex,
so that $q_1 +q_3 =l'_1 -l_1 =\lambda$.
The other target (cut line) has the momentum $l_2$, it interacts
with the reggeon $q_2$ on the right side of the cut
and with the reggeon $q'_2=q-p$ on the left side
(Diagram Fig.~\ref{FC},1 presents the momentum flow).
From the conservation law
$q'_2=q-p=q_2 +l_2 -l'_2 =q_2 +\lambda$
the relation $q'_{2\perp}=q_{2\perp}$
follows (used in \Ref{e11}).
Reggeons with momenta $q_1$, $q_2$, $q_3$ are labeled as
1,2,3 in Figs.~\ref{C1}-\ref{A1}, the correspondent
color indices are denoted as $b_1$, $b_2$, $b_3$, respectively.

To invoke the overall colorless exchange we assume that the
color structure of the diagrams corresponds to interaction with
quark loops. Closing of quark loops also implied for the spinor
structure, although we do not perform loop momentum integration.
Thus for all diagrams we have the color factor for the  target
quark lines (both cut and uncut)
$\tr(T^{b} T^{b'})/N_c=\delta^{bb'}/2N_c$.
Here factor $1/N_c$ compensates the sum over $N_c$ colors.
The rest color factors are different and will be calculated
separately for each diagram.

For the spinor factors we sum over spins
of intermediate quark states
and take an average over spins of the initial and the final quark states.
So for the projectile quark we have (here and below we use the Regge
approximation neglecting small components of momenta)
\begin{equation}
\frac{1}{2}\sum_{\sigma} \bar{u}_{\sigma}({\bf k}) \gamma_{+}
\cdot(\hat{k}-\hat{q})\cdot \gamma_{+} u_{\sigma}({\bf k})
= \frac{1}{2} \tr(\gamma_{+}(\hat{k}-\hat{q})\gamma_{+}\hat{k})
\approx 4k_{+}^2 \ ,
\label{e12}
\end{equation}
for the uncut target quark line:
\begin{equation}
\frac{1}{2}\sum_{\sigma} \bar{u}_{\sigma}({\bf l'_1})
\frac{i\gamma_{-}}{2}
\frac{\hat l_1 + \hat q_1}{l_{1-} +  q_{1-}}
\frac{i\gamma_{-}}{2} u_{\sigma}({\bf l_1})
\approx  -l_{1-}  = -l_{-}
\label{e13}
\end{equation}
and for the cut one:
\begin{equation}
\frac{1}{2}\sum_{\sigma} \bar{u}_{\sigma}({\bf l'_2})
\gamma_{-} \cdot(\hat l_2 + \hat q_2)\cdot
\frac{i\gamma_{-}}{2} u_{\sigma}({\bf l_2})
\approx  2i l_{2-}^2 = 2i l_{-}^2 \ .
\label{e14}
\end{equation}
These factors are common for all diagrams of this Section.
The common order in the coupling constant is $g^{10}$.

\subsection{Calculation of integrals over longitudinal variables}

With our notation,
momenta $q_1$ and $q_2$ are independent and can be chosen as
variables of integration. Using the conservation law $q=p+q_2+\lambda$
and mass-shell conditions for cut quark lines we can integrate
over two of four longitudinal components
(factors $1/2$ are from our normalization of longitudinal components
$q_{\pm}=q_0 \pm q_z$):
$$
\int\frac{d^4 q_1}{(2\pi)^4}\int\frac{d^4 q_2}{(2\pi)^4}
2\pi\delta((k-q)^2)\cdot
2\pi\delta((l_2+q_2)^2)
$$
\begin{equation}
=
\frac{1}{2k_+ l_{2-}}
\int\frac{dq_{1+} dq_{1-}}{2(2\pi)^2}
\int\frac{d^2 q_{1\perp} d^2 q_{2\perp}}{(2\pi)^4} \ .
\label{cr12}
\end{equation}
Due to the Regge kinematics \Ref{e1} in what follows
we have to neglect momentum components
\begin{equation}
q_-= p_- +q_{2-} +\lambda_- = -\frac{q_\perp^2}{k_+-q_+}
\label{cr13}
\end{equation}
and
\begin{equation}
q_{2+}=-l_{2+}-\frac{q_{2\perp}^2}{l_{2-}+q_{2-}} \ .
\label{cr14}
\end{equation}

Calculation of integrals over the remaining longitudinal components
can be carried out in a general form.
For all diagrams considered in this and the following sections
with our notation of reggeon momenta
the momentum of the virtual target quark interacting with
two reggeons is always $l_1 + q_1$.
As argued in \cite{BLSV}, the propagator of the fast moving
target quark in the Lipatov effective theory is not
\begin{equation}
i \frac{\hat{l_1}+\hat{q_1}} {(l_1 + q_1)^2 +i0}\ ,
\label{e20}
\end{equation}
but is reduced to its delta-functional part
\begin{equation}
i(\hat{l_1}+\hat{q_1})\  (-i\pi) \delta((l_1 + q_1)^2)
=
\frac{\hat{l_1}+\hat{q_1}}{l_{1-}+q_{1-}}\ \pi\delta(q_{1+})
%\!\left(q_{1+}+l_{1+}+\frac{(l_1 + q_1)_{\perp}^2}{l_{1-}+q_{1-}}\right).
\label{e21}
\end{equation}
where we used $l_{1+}=0$ and $l_{1-}>>|q_{1\perp}|$.
Then the common integral with two propagators has the form
$$
I_2(p_1)=
\int\!\frac{dq_{1-}dq_{1+}}{2(2\pi)^2}
\frac{1}{(p_1 - q_1)^2 +i0}
\ \pi\delta(q_{1+})
$$
\begin{equation}
=\frac{1}{4}
\int\!\frac{dq_{1-}}{2\pi}
\frac{1}
{(p_{1-}-q_{1-})p_{1+}
+(p_1 - q_1)_{\perp}^2 +i0 } \ .
\label{e23}
\end{equation}
Neglecting the transverse part and splitting the integrand into the
principal value part and $\delta$-function we get
\begin{equation}
I_2(p_1) \approx \frac{-i}{8|p_{1+}|} \ .
\label{e25}
\end{equation}

The common integral with three propagators has the general form
$$
I_3(p_1,p_2)=
\int\!\frac{dq_{1-}dq_{1+}}{2(2\pi)^2}
\frac{1}{((p_1 - q_1)^2 +i0)((p_2 + q_1)^2 +i0)}
\ \pi
\delta(q_{1+})
$$
\begin{equation}
\approx
\frac{-i}{4 p_{1+} p_{2+}} \,\cdot\,
\frac{{\rm sign}(p_{1+})}
{p_{1-}+p_{2-}
+i0 \cdot {\rm sign}(p_{1+})} \ ,
\label{e26}
\end{equation}
when $p_{1+}p_{2+}>0$, and turns to zero,
when $p_{1+}p_{2+}<0$.

After integration over longitudinal components
we are left with the integral over transverse components
\begin{equation}
\int\frac{d^2 q_{1\perp} d^2 q_{2\perp}}{(2\pi)^4}
\frac{1}{(q_{1\perp}^2 q_{2\perp}^2)^2} \ .
\label{e27}
\end{equation}
This integration is always implied in final results and
will be suppressed in the following
together with the common factor
\begin{equation}
(-i)8g^{10}\frac{4k_{+}^2 (-l_{1-}) 2i l_{2-}^2} {2k_+ l_{2-}} L(p,q_2)
=-32 g^{10} k_{+} l_{1-} l_{2-} L(p,q_2)
=-32 g^{10} s\sqrt{s} L(p,q_2) \ .
\label{e28}
\end{equation}

\begin{figure}
\begin{center}
\includegraphics[scale=0.55]{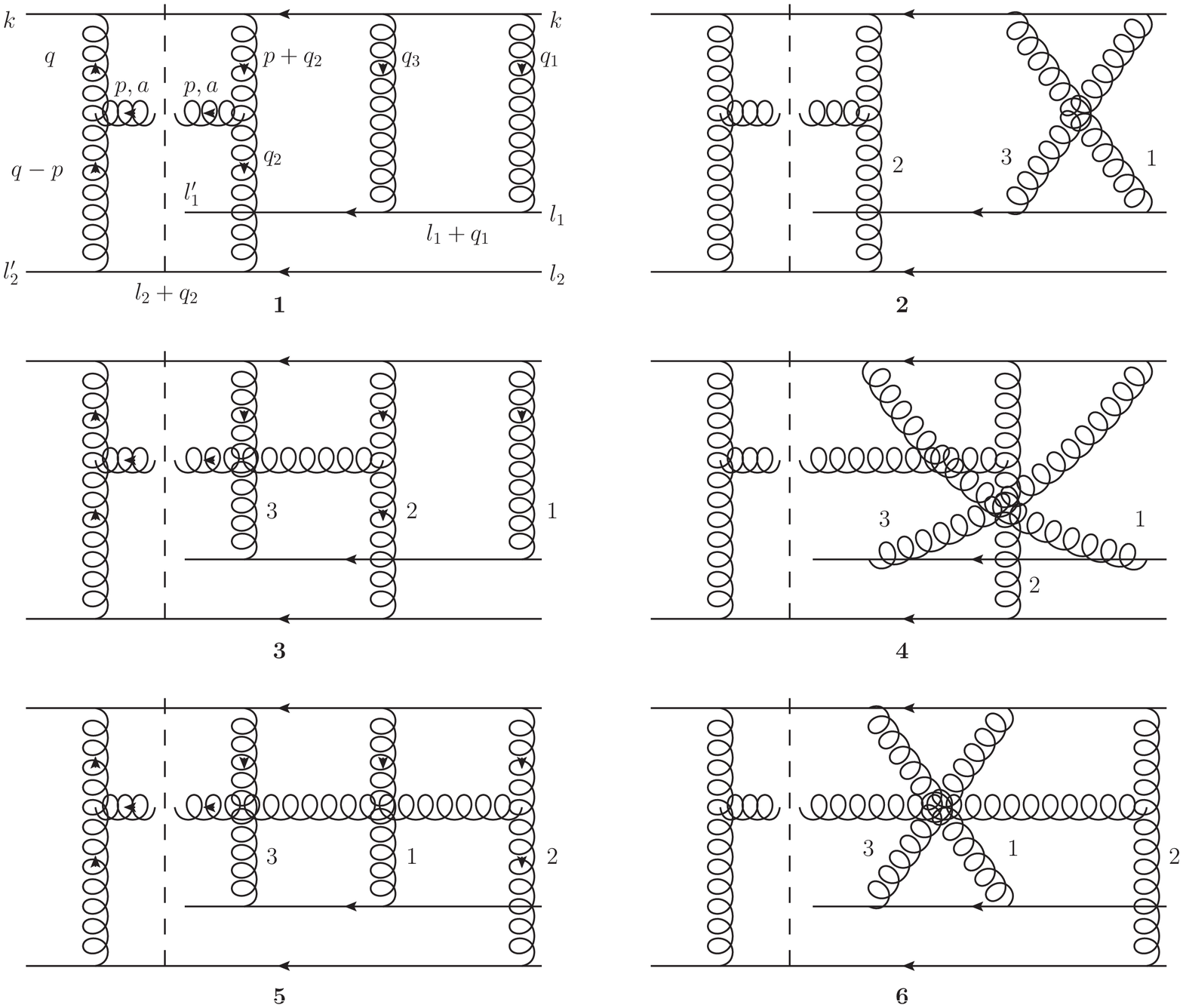}
\includegraphics[scale=0.55]{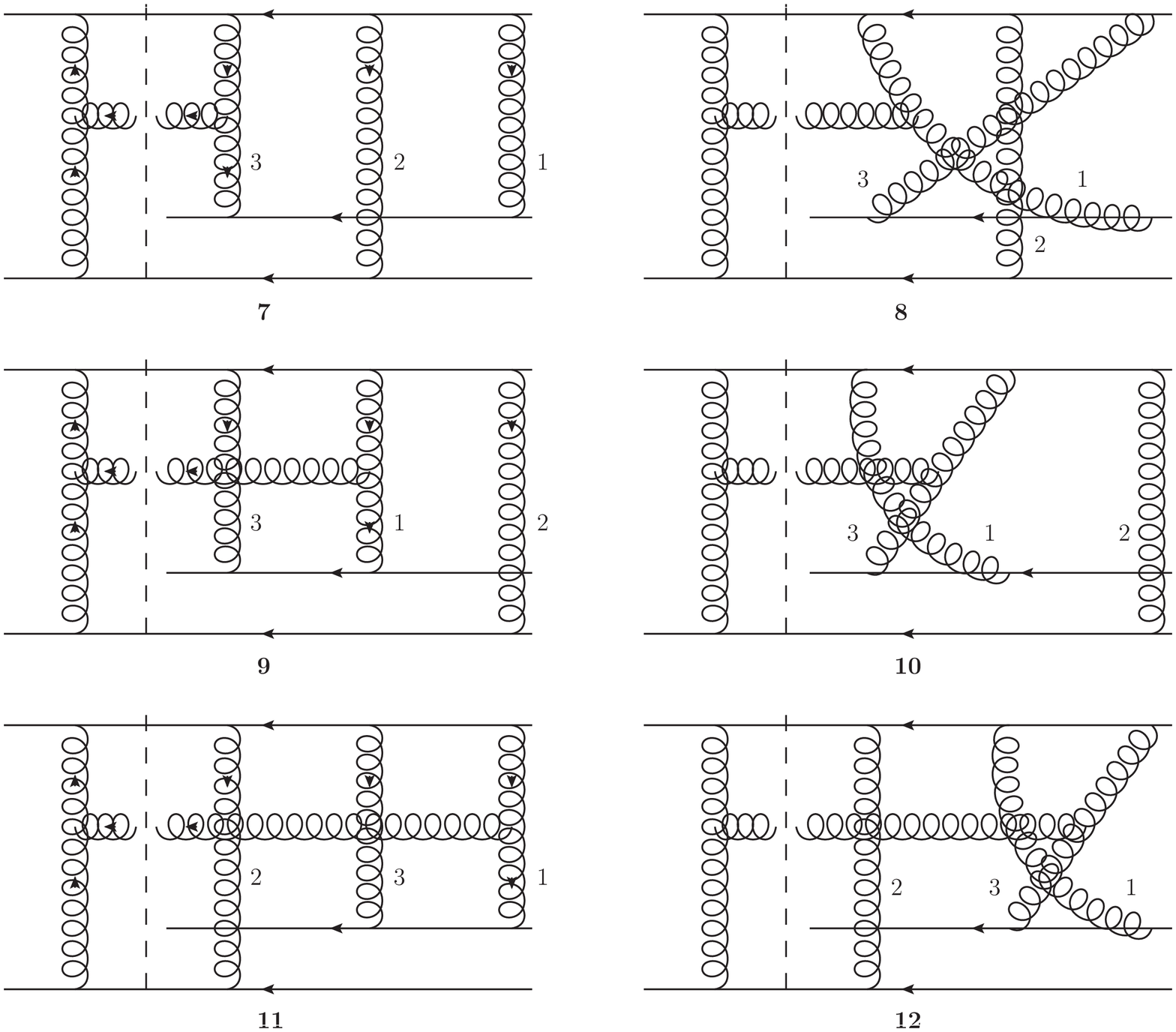}
\end{center}
\end{figure}
\begin{figure}[h]
\begin{center}
\includegraphics[scale=0.55]{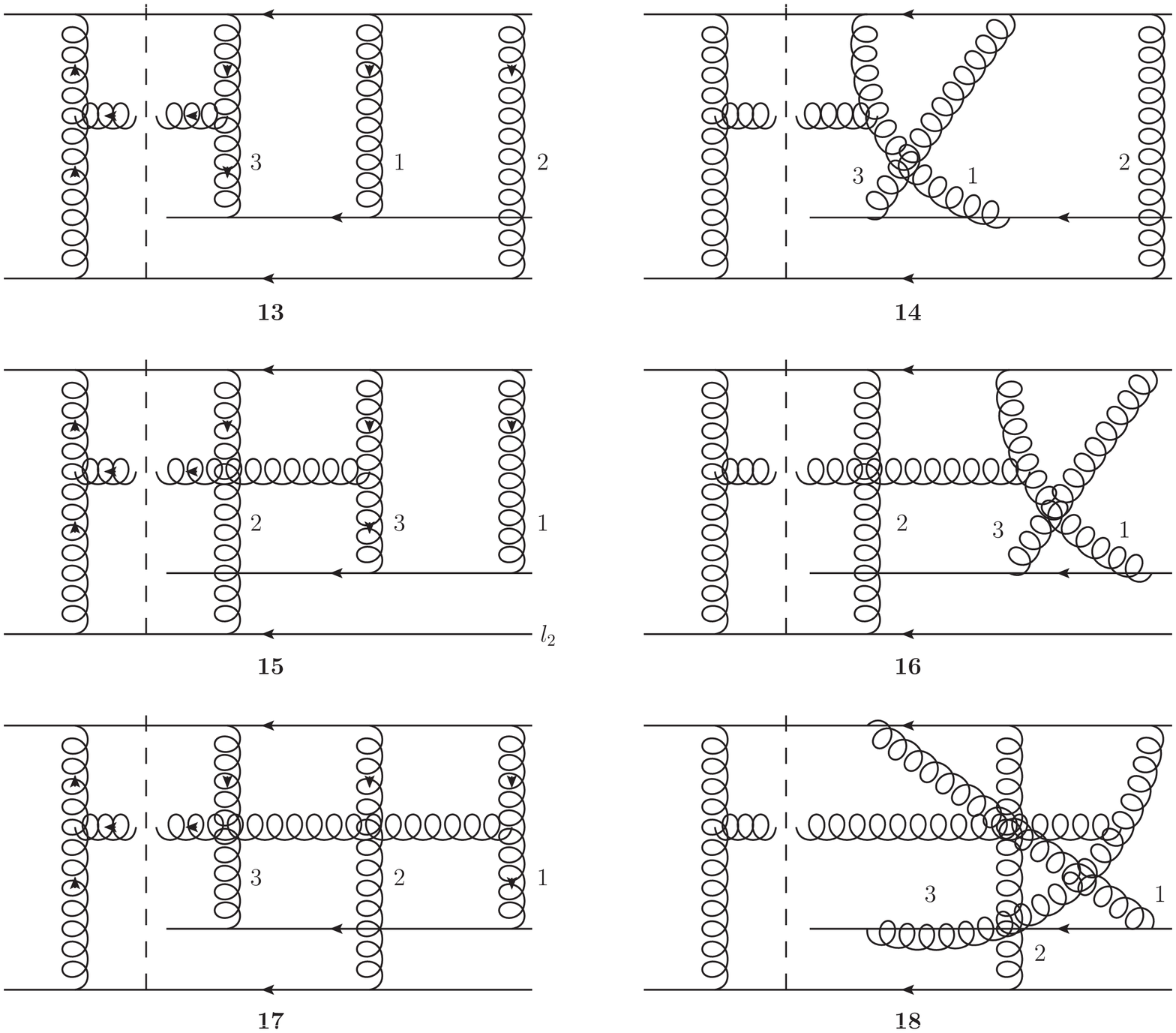}
\end{center}
\caption{Diagrams with R$\to$RP vertices.}
\label{C1}
\end{figure}

\subsection{Calculation of diagrams from Fig.~\ref{C1}}

Diagrams with gluon production from the R$\to$RP (Lipatov) vertex,
corresponding  to the part \Ref{ea3}
of the amplitude, are shown in Fig.~\ref{C1}.

For the momentum factor of the  diagram 1 from Fig.~\ref{C1}
we get
\begin{equation}
\frac{k_+^2 I_2(k) L(p,q_2)}{(k-q_1 -q_3)^2+i0}
=
i\frac{L(p,q_2)}{8(\lambda_{-} -i0)} \ .
\label{emc1}
\end{equation}
The momentum factor for diagram 2 from Fig.~\ref{C1} is
\begin{equation}
\frac{k_+^2 I_2(\lambda-k) L(p,q_2)}{(k-\lambda)^2+i0}
=
i\frac{L(p,q_2)}{8(\lambda_{-} -i0)} \ .
\label{emc2}
\end{equation}

The color structure of diagram 1 from Fig.~\ref{C1} is
\begin{equation}
\frac{1}{4N_c^3}f^{b' b_2 a} \cdot
\tr(T^{b'} T^d T^{b_1} T^{b_1}) f^{d b_2 a}
=
\frac{1}{4N_c^2}\tr(T^{b'} T^{b'} T^{b_1} T^{b_1})
=
\frac{(N_c^2 -1)^2}{16N_c^3} \ .
\label{ecc1}
\end{equation}
Diagram 2 differs from diagram 1 only by the exchange of reggeons 1 and 3.
Since these reggeons with color indices $b_1$ and $b_3$ form a symmetrical
colorless state, the color factor for Diagram 2 from Fig.~\ref{C1}
is the same as for diagram 1.

For the rest diagrams we present a list of results
where diagrams are arranged in pairs having a common color factor.

{\bf 5,6.} The momentum factor for diagram 5 from Fig.~\ref{C1} is
$$
\frac{k_+^2 I_2(k-q+\lambda) L(p,q_2)}{(k-q+\lambda)^2+i0}
=
-i\frac{L(p,q_2)}{8(\lambda_{-} +i0)}
$$
and for diagram 6 is
$$
\frac{k_+^2 I_2(q-k) L(p,q_2)}{(k-q+\lambda)^2+i0}
=
-i\frac{L(p,q_2)}{8(\lambda_{-} +i0)} \ ,
$$
the common color factor is
$$
\frac{1}{4N_c^3}f^{b' b_2 a} \cdot
\tr(T^{b'} T^{b_1} T^{b_1} T^d) f^{d b_2 a}
=\frac{(N_c^2 -1)^2}{16N_c^3} \ .
$$

{\bf 9,10.} The momentum factor for diagram 9 from Fig.~\ref{C1} is
$$
\frac{k_+^2 I_2(k-q+\lambda) L(p,q_1)}{(k-q+p+\lambda)^2+i0}
=
-i\frac{L(p,q_1)}{8(\lambda_{-} +i0)}
$$
and for diagram 10 is
$$
\frac{k_+^2 I_2(q-k) L(p,q_3)}{(k-q+p+\lambda)^2+i0}
=
-i\frac{L(p,q_3)}{8(\lambda_{-} +i0)} \ ,
$$
the common color factor is
$$
\frac{1}{4N_c^3}f^{b' b_2 a} \cdot
\tr(T^{b'} T^{b_1} T^d T^{b_2}) f^{d b_1 a}
=-\frac{N_c^2 -1}{32N_c} \ .
$$

{\bf 11,12.} The momentum factor for diagram 11 from Fig.~\ref{C1} is
$$
\frac{k_+^2 I_2(k-p) L(p,q_1)}{(k-p-\lambda)^2+i0}
=
i\frac{L(p,q_1)}{8(\lambda_{-} -i0)}
$$
and for diagram 12 is
$$
\frac{k_+^2 I_2(p-k+\lambda) L(p,q_3)}{(k-p-\lambda)^2+i0}
=
i\frac{L(p,q_3)}{8(\lambda_{-} -i0)} \ ,
$$
the common color factor is
$$
\frac{1}{4N_c^3}f^{b' b_2 a} \cdot
\tr(T^{b'} T^{b_2} T^{b_1} T^d) f^{d b_1 a}
=\frac{N_c^2 -1}{32N_c} \ .
$$

{\bf 13,14.} The momentum factor for diagram 13 from Fig.~\ref{C1} is
$$
\frac{k_+^2 I_2(k-q+p+\lambda) L(p,q_3)}{(k-q+p+\lambda)^2+i0}
=
-i\frac{L(p,q_3)}{8(\lambda_{-} +i0)}
$$
and for diagram 14 is
$$
\frac{k_+^2 I_2(q-k-p) L(p,q_1)}{(k-q+p+\lambda)^2+i0}
=
-i\frac{L(p,q_1)}{8(\lambda_{-} +i0)} \ ,
$$
the common color factor is
$$
\frac{1}{4N_c^3}f^{b' b_2 a} \cdot
\tr(T^{b'} T^d T^{b_1} T^{b_2}) f^{d b_1 a}
=\frac{N_c^2 -1}{32N_c} \ .
$$

{\bf 15,16.} The momentum factor for diagram 15 from Fig.~\ref{C1} is
$$
\frac{k_+^2 I_2(k) L(p,q_3)}{(k-p-\lambda)^2+i0}
=
i\frac{L(p,q_3)}{8(\lambda_{-} -i0)}
$$
and for diagram 16 is
$$
\frac{k_+^2 I_2(\lambda-k) L(p,q_1)}{(k-p-\lambda)^2+i0}
=
i\frac{L(p,q_1)}{8(\lambda_{-} -i0)} \ ,
$$
the common color factor is
$$
\frac{1}{4N_c^3}f^{b' b_2 a} \cdot
\tr(T^{b'} T^{b_2} T^d T^{b_1}) f^{d b_1 a}
=-\frac{N_c^2 -1}{32N_c} \ .
$$

{\bf 3,4,7,8,17,18.} The momentum factor for diagram 3 from Fig.~\ref{C1} is
$$
k_+^2 I_3(k, q-k-\lambda) L(p,q_2)=0,
$$
because the sign of $k_{+}(q-k-\lambda)_{+} = -k_{+}^2$ is negative.
For the same reason the momentum factors are zero for diagrams
4,7,8,17,18 from Fig.~\ref{C1}.

Taking into account the common ''-'' sign from \Ref{ea3} we find
the total contribution from diagrams 1-6 from Fig.~\ref{C1}:
$$
-\frac{(N_c^2 -1)^2}{16N_c^3}
\left( 2i\frac{L(p,q_2)}{8(\lambda_{-} -i0)}
-2i\frac{L(p,q_2)}{8(\lambda_{-} +i0)}\right)
=
\frac{(N_c^2 -1)^2}{32N_c^3}
L(p,q_2)\ \pi\delta(\lambda_{-})
$$
\begin{equation}
=
\frac{(N_c^2 -1)^2}{64N_c^3}
L(p,q_2)\sqrt{s}\ 2\pi\delta(2k\lambda) ,
\label{e3c1}
\end{equation}
whereas contributions from diagrams 7-18 cancel completely.
Taking into account the complex conjugate term doubles
this result.

The ''double cut'' contribution from diagrams such as in Fig.~\ref{npap5}
to the coefficient $F$ of the singularity $F\delta(\lambda_z)$
of the discontinuity,
where $\lambda_z$ is the $z$-component in the lab. system,
was found in \cite{BSV}
as a sum of four terms $F=F_1 +F_2 +F_3 +F_4$.
Now we prefer to define the singular part as
(see Eq. \Ref{handf1} in Appendix 1)
\begin{equation}
2\pi \delta(2k\lambda) F
=\frac{2\pi}{k_+}\delta(\lambda_-^{c.m.}) F
=\frac{2\pi m}{s}\delta(\lambda_z^{lab.sys.}) F ,
\label{e3n}
\end{equation}
therefore we have to divide the old result by $2\pi m/s$.
Also the overall coefficient has to be taken two times smaller
than in \cite{BSV} since the discontinuity is twice
the imaginary part.
Then the previously found contributions can be presented
in the following form:
$$
F_1= s^2 g^{10} \frac{N_c^2-1}{N_c}
 B^2(p,q_2,q_1) ,
\quad
F_2= s^2 g^{10} \frac{(N_c^2-1)^2}{N_c^3}
 L^2(p,q_1) ,
$$
\begin{equation}
F_3= -\frac{1}{2} s^2 g^{10} \frac{N_c^2-1}{N_c}
 L(p,q_1)L(p,q_2),
\quad
F_4= -\frac{1}{2} s^2 g^{10} \frac{N_c^2-1}{N_c}
 B(p,q_2,q_1) \Big( L(p,q_1) - 2L(p,q_2) \Big) ,
\label{e3o}
\end{equation}
where the transverse integration \Ref{e27} is implied.

Analogously, restoring the common factor \Ref{e28}
we find the new contribution to the coefficient $F$
of the imaginary part:
\begin{equation}
F_5=-s^2 g^{10}\frac{(N_c^2-1)^2}{N_c^3}
L^2(p,q_2).
\label{e3f5}
\end{equation}

\begin{figure}
\begin{center}
\includegraphics[scale=0.43]{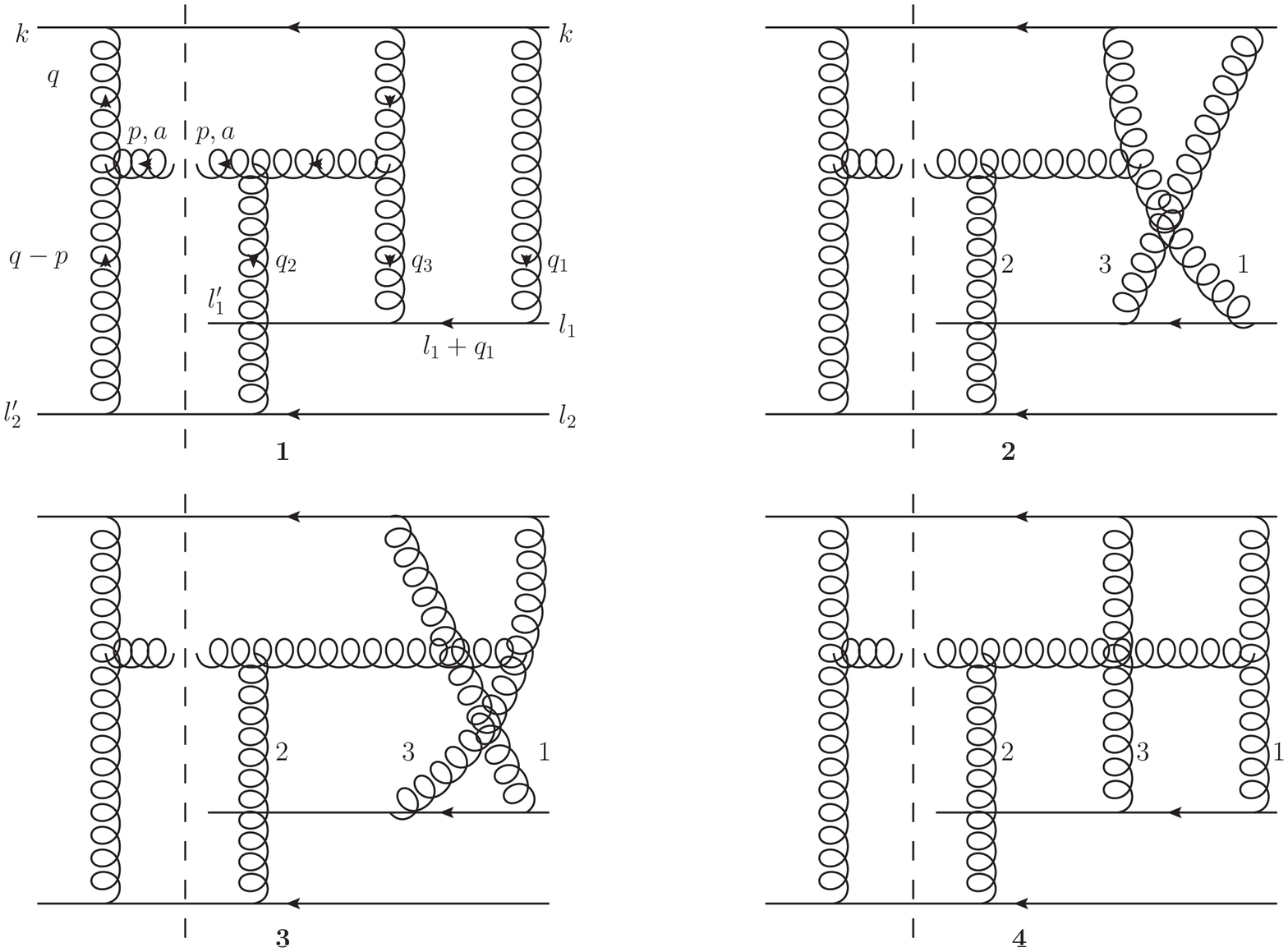}
\includegraphics[scale=0.43]{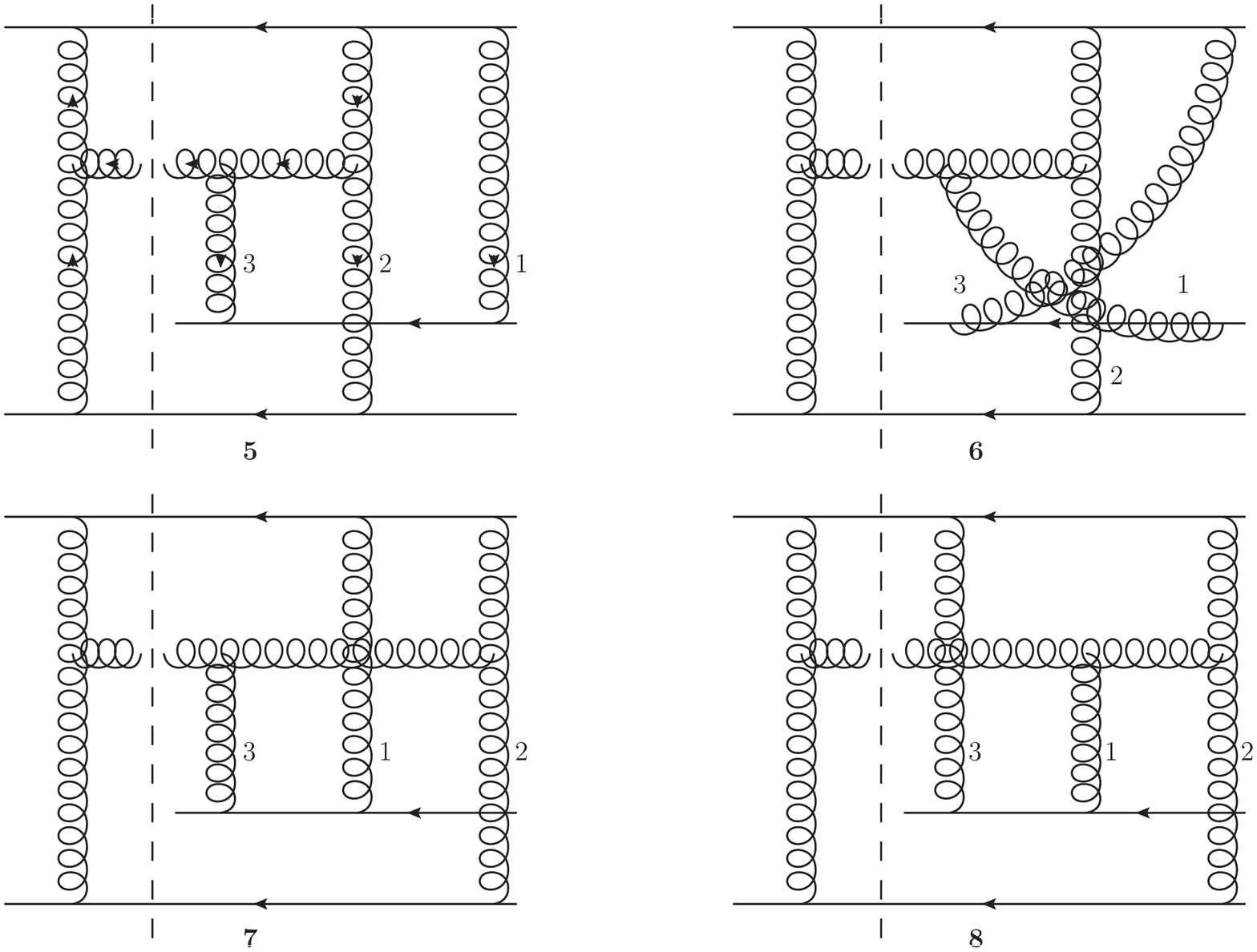}
\includegraphics[scale=0.43]{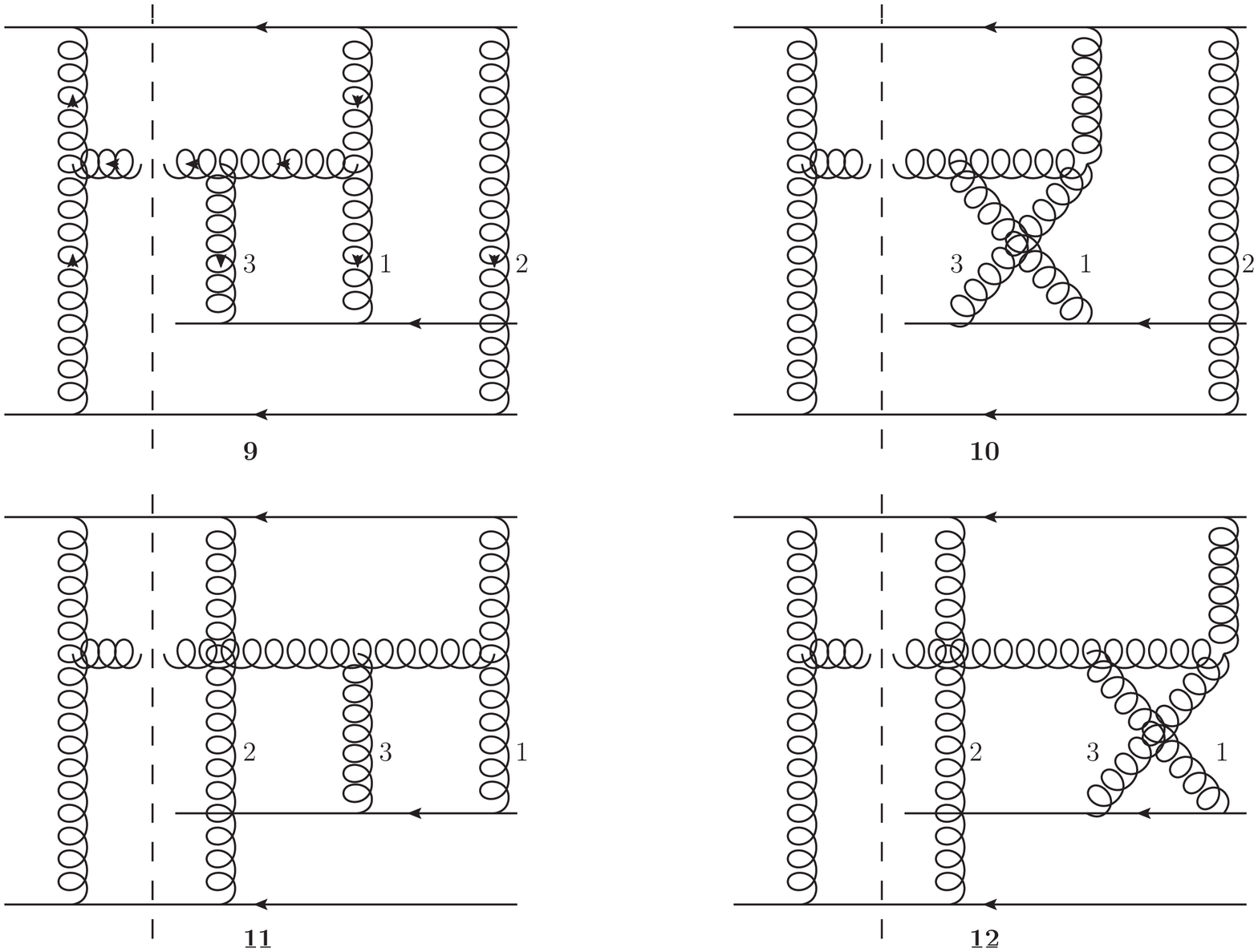}
\end{center}
\caption{Diagrams with R$\to$RRP vertices.}
\label{B1}
\end{figure}

\subsection{Calculation of diagrams from Fig.~\ref{B1}}

Diagrams with gluon production from R$\to$RRP (Bartels) vertex,
corresponding to the part  \Ref{ea2}
of the amplitude, are shown in Fig.~\ref{B1}.
Color factors are also coincide for each pair of diagrams
which differ only by the exchange of reggeons 1 and 3.

{\bf 1,2.} The momentum factor for diagram 1 from Fig.~\ref{B1} is
$$
\frac{k_+ q_+ I_2(k) B(p,q_2,q_3)}{(q-\lambda)^2+i0}
=
i\frac{B(p,q_2,q_3)}{8(\lambda_{-} -i0)}
$$
and for diagram 2 is
$$
\frac{k_+ q_+ I_2(\lambda -k) B(p,q_2,q_1)}{(q-\lambda)^2+i0}
=
i\frac{B(p,q_2,q_1)}{8(\lambda_{-} -i0)} \ ,
$$
the common color factor is
$$
\frac{i}{4N_c^3}f^{b' b_2 a} \cdot
\tr(T^{b'} T^c T^{b_1}) f^{c b_1 d} f^{d b_2 a}
=-\frac{N_c^2 -1}{16N_c} \ .
$$

{\bf 3,4.} The momentum factor for diagram 3 from Fig.~\ref{B1} is
$$
\frac{k_+ q_+ I_2(q-k) B(p,q_2,q_3)}{(q-\lambda)^2+i0}
=
i\frac{B(p,q_2,q_3)}{8(\lambda_{-} -i0)}
$$
and for diagram 4 is
$$
\frac{k_+ q_+ I_2(k-q+\lambda) B(p,q_2,q_1)}{(q-\lambda)^2+i0}
=
i\frac{B(p,q_2,q_1)}{8(\lambda_{-} -i0)} \ ,
$$
the common color factor is
$$
\frac{i}{4N_c^3}f^{b' b_2 a} \cdot
\tr(T^{b'} T^{b_1} T^c) f^{c b_1 d} f^{d b_2 a}
=\frac{N_c^2 -1}{16N_c} \ .
$$

{\bf 5,6.} The momentum factor for diagram 5 from Fig.~\ref{B1} is
$$
k_+ q_+ I_3(k, -p-\lambda) B(p,q_3,q_2)=0,
$$
because the sign of $k_{+}(-p-\lambda)_{+} = -k_{+}p_{+}$ is negative,
and also for Diagram 6 is
$$
k_+ q_+ I_3(\lambda-k, p) B(p,q_1,q_2)=0.
$$

{\bf 7,8.} The momentum factor for diagram 7 from Fig.~\ref{B1} is
$$
k_+ q_+ I_3(q-k, -p-\lambda) B(p,q_3,q_2)
=
-i\frac{B(p,q_3,q_2)}{4(\lambda_{-} +i0)}
$$
where we have taken into account that $k_+>>p_+$.
For diagram 8  the momentum factor is
$$
k_+ q_+ I_3(k-q+\lambda, p) B(p,q_1,q_2)
=
-i\frac{B(p,q_1,q_2)}{4(\lambda_{-} +i0)} \ .
$$
The common color factor is
$$
\frac{i}{4N_c^3}f^{b' b_2 a} \cdot
\tr(T^{b'} T^{b_1} T^c) f^{c b_2 d} f^{d b_1 a}
=\frac{N_c^2 -1}{32N_c} \ .
$$

{\bf 9,10.} The momentum factor for diagram 9 from Fig.~\ref{B1} is
$$
\frac{k_+ q_+ I_2(p+\lambda) B(p,q_3,q_1)}{(k-q+p+\lambda)^2+i0}
=
-i\frac{B(p,q_3,q_1)}{8(\lambda_{-} +i0)}
$$
and for diagram 10 is
$$
\frac{k_+ q_+ I_2(-p) B(p,q_1,q_3)}{(k-q+p+\lambda)^2+i0}
=
-i\frac{B(p,q_1,q_3)}{8(\lambda_{-} +i0)} \ ,
$$
the common color factor is
$$
\frac{i}{4N_c^3}f^{b' b_2 a} \cdot
\tr(T^{b'} T^c T^{b_2}) f^{c b_1 d} f^{d b_1 a}
=-\frac{N_c^2 -1}{16N_c} \ .
$$

{\bf 11,12.} The momentum factor for diagram 11 from Fig.~\ref{B1} is
$$
\frac{k_+ q_+ I_2(p+\lambda) B(p,q_3,q_1)}{(k-p-\lambda)^2+i0}
=
i\frac{B(p,q_3,q_1)}{8(\lambda_{-} -i0)}
$$
and for diagram 12 is
$$
\frac{k_+ q_+ I_2(-p) B(p,q_1,q_3)}{(k-p-\lambda)^2+i0}
=
i\frac{B(p,q_1,q_3)}{8(\lambda_{-} -i0)} \ ,
$$
the common color factor is
$$
\frac{i}{4N_c^3}f^{b' b_2 a} \cdot
\tr(T^{b'} T^{b_2} T^c) f^{c b_1 d} f^{d b_1 a}
=\frac{N_c^2 -1}{16N_c} \ .
$$

The total sum is
$$
\frac{i}{8}\cdot\frac{N_c^2 -1}{16N_c} \left(
-\frac{B(p,q_3,q_2)}{\lambda_{-} +i0}
-\frac{B(p,q_1,q_2)}{\lambda_{-} +i0}
+\frac{B(p,q_3,q_1)}{\lambda_{-} +i0}
+\frac{B(p,q_1,q_3)}{\lambda_{-} +i0}
+\frac{B(p,q_3,q_1)}{\lambda_{-} -i0}
+\frac{B(p,q_1,q_3)}{\lambda_{-} -i0}
\right)
$$
\begin{equation}
=\frac{i}{8}\cdot\frac{N_c^2 -1}{16N_c} \left(
-\frac{B(p,q_3,q_2)+B(p,q_1,q_2)}{\lambda_{-} +i0}
+2\left(B(p,q_3,q_1)+B(p,q_1,q_3)\right)
\cdot P\frac{1}{\lambda_{-}} \right) ,
\label{e3b1}
\end{equation}
where only diagrams 7-12 contribute.
Taking into account the complex conjugate term leaves us with
\begin{equation}
-\frac{N_c^2 -1}{32N_c}\cdot
\frac{B(p,q_3,q_2)+B(p,q_1,q_2)}{2}
\pi\delta(\lambda_{-})
=
-\frac{N_c^2 -1}{64N_c}\cdot
\frac{B(p,q_3,q_2)+B(p,q_1,q_2)}{2}\sqrt{s}
\ 2\pi\delta(2k\lambda) .
\label{e3b2}
\end{equation}
Using the invariance of the transverse integration \Ref{e27}
with respect to changing the variable
$q_{1\perp}\to q_{3\perp}= \lambda_{\perp}-q_{1\perp}=-q_{1\perp}$
and restoring the common factor
we find the contribution of diagrams from Fig.~\ref{B1}
to the coefficient $F$:
\begin{equation}
F_6=\frac{1}{2} s^2 g^{10}\frac{N_c^2-1}{N_c}
L(p,q_2)B(p,q_1,q_2) .
\label{e3f6}
\end{equation}

\begin{figure}[h]
\begin{center}
\includegraphics[scale=0.55]{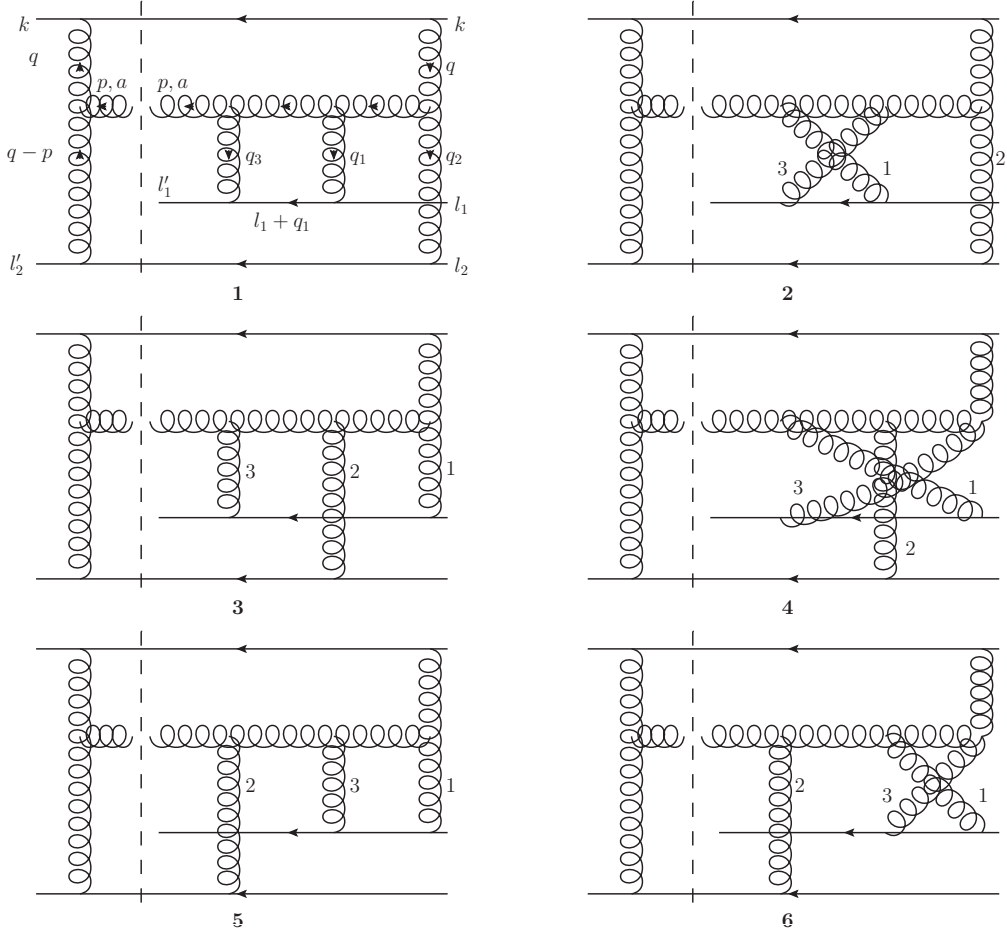}
\end{center}
\caption{Diagrams with R$\to$RRRP vertices.}
\label{A1}
\end{figure}

\subsection{Calculation of diagrams from Fig.~\ref{A1}}

Diagrams with gluon production from R$\to$RRRP vertex,
corresponding to the part \Ref{ea1}
of the amplitude, are shown in Fig.~\ref{A1}.

{\bf 1,2.} The momentum factor for diagram 1 from Fig.~\ref{A1} is
$$
\frac{q_+^2 I_2(p+\lambda) B(p,q_3 +q_1,q_2)}{(p+\lambda)^2+i0}
=
-i\frac{B(p,q_3 +q_1,q_2)}{8(\lambda_{-} +i0)}
$$
and for diagram 2 is
$$
\frac{q_+^2 I_2(-p) B(p,q_1 +q_3,q_2)}{(p+\lambda)^2+i0}
=
-i\frac{B(p,q_3 +q_1,q_2)}{8(\lambda_{-} +i0)}\ .
$$
The common color factor is
$$
\frac{1}{4N_c^3}f^{b' b_2 a} \cdot
\tr(T^{b'} T^b) f^{b b_2 c} f^{c b_1 d} f^{d b_1 a}
=-\frac{N_c^2 -1}{8N_c} \ .
$$

{\bf 3,4.} The momentum factor for diagram 3 from Fig.~\ref{A1} is
$$
q_+^2 I_3(q,-p-\lambda) B(p,q_3 +q_2,q_1)=0,
$$
because the sign of $q_{+}(-p-\lambda)_{+} = -q_{+}^2$ is negative.
Also for diagram 4 the momentum factor is zero:
$$
q_+^2 I_3(\lambda-q,p) B(p,q_1 +q_2,q_3)=0.
$$

{\bf 5,6.} The momentum factor for diagram 5 from Fig.~\ref{A1} is
$$
\frac{q_+^2 I_2(p+\lambda) B(p,q_2 +q_3,q_1)}{(q-\lambda)^2+i0}
=
i\frac{B(p,q_2 +q_3,q_1)}{8(\lambda_{-} -i0)}
$$
and for diagram 6 is
$$
\frac{q_+^2 I_2(-p) B(p,q_2 +q_1,q_3)}{(p+\lambda)^2+i0}
=
i\frac{B(p,q_2 +q_1,q_3)}{8(\lambda_{-} -i0)} \ .
$$
The common color factor is
$$
\frac{1}{4N_c^3}f^{b' b_2 a} \cdot
\tr(T^{b'} T^b) f^{b b_1 c} f^{c b_1 d} f^{d b_2 a}
=-\frac{N_c^2 -1}{8N_c} \ .
$$

To calculate the total contribution we take into account that
\begin{equation}
B(p,q_3 +q_1,q_2)=
\frac{(p+q_3 +q_1,\epsilon_\perp)}{(p+q_3 +q_1)_\perp^2}
- \frac{(p+q_3+q_1+q_2,\epsilon_\perp)}{(p+q_3+q_1+q_2)_\perp^2}
\equiv
L(p,q_2)
\label{e3a1}
\end{equation}
and
$$
B(p,q_2 +q_1,q_3)=
\frac{(p+q_2 +q_1,\epsilon_\perp)}{(p+q_2 +q_1)_\perp^2}
- \frac{(p+q_2+q_1+q_3,\epsilon_\perp)}{(p+q_2+q_1+q_3)_\perp^2}
$$
$$
=
\frac{(p+q_2 +q_1,\epsilon_\perp)}{(p+q_2 +q_1)_\perp^2}
- \frac{(p+q_2,\epsilon_\perp)}{(p+q_2)_\perp^2}
\equiv
-B(p,q_2,q_1),
$$
\begin{equation}
B(p,q_2 +q_3,q_1)
\equiv
-B(p,q_2,q_3),
\label{e3a2}
\end{equation}
since $q_{1\perp}+q_{3\perp}=\lambda_{\perp}=0$.
Then we get together with the complex conjugate term
$$
-\frac{N_c^2 -1}{8N_c} \left(
-2i\frac{L(p,q_2)}{8(\lambda_{-} +i0)}
+i\frac{-B(p,q_2,q_1)-B(p,q_2,q_3)}{8(\lambda_{-} -i0)}
\right) + c.c.
$$
\begin{equation}
=\frac{N_c^2 -1}{16N_c} \left(
L(p,q_2)-B(p,q_2,q_1)
\right) \pi\delta(\lambda_{-})
=
\frac{N_c^2 -1}{32N_c} \left(
L(p,q_2)-B(p,q_2,q_1) \right)\sqrt{s}
\ 2\pi\delta(2k\lambda) ,
\label{e3a3}
\end{equation}
where the invariance with respect to the change
$q_{1\perp}\lra q_{3\perp}$ under the sign of integration
\Ref{e27} is used.
Restoring the common factor \Ref{e28} we find the contribution
of diagrams from Fig.~\ref{A1} to the coefficient $F$
as the sum of two terms:
$$
F_7=-s^2 g^{10}\frac{N_c^2-1}{N_c}
L^2(p,q_2)
\quad\quad \mbox{and}
$$
\begin{equation}
F_8=s^2 g^{10}\frac{N_c^2-1}{N_c}
L(p,q_2)B(p,q_2,q_1) .
\label{e3f78}
\end{equation}

\begin{figure}[h]
\begin{center}
\includegraphics[scale=0.60]{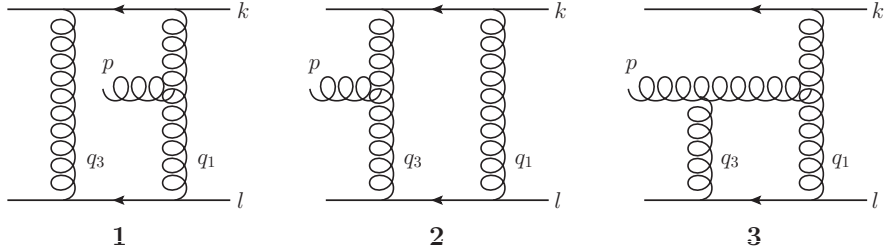}
\end{center}
\caption{Production amplitude in the diffractive configuration.}
\label{D1}
\end{figure}

\section{Diffractive contribution}

In this section we calculate the contribution from the
diffractive configuration.
The corresponding diagrams  are shown
in Fig~\ref{D1}. Three more diagrams with exchanged reggeons
of momenta $q_1$ and $q_3$ have to be added. Here we denote reggeon
momenta $q_1$ and $q_3$ to preserve the kinematical
relation $q_1+q_3=\lambda$ with $q_{1\perp}+q_{3\perp}=0$.

The initial formulae for the calculation of the amplitude
were obtained in  \cite{BV,BLSV}.
For all diagrams from Fig.~\ref{D1} the propagator \Ref{e21}
of the target quark multiplied by two reggeon propagators
and two reggeon-quark vertices is
\begin{equation}
\frac{i\gamma_{-}}{2} \left[
i(\hat l_1 + \hat q_1) (-i\pi) \delta((l_1 + q_1)^2) \right]
\frac{i\gamma_{-}}{2}
\cdot \frac{-2i}{q_{1\perp}^2} \cdot \frac{-2i}{q_{3\perp}^2}
=
\frac{\gamma_{-}}{(q_{1\perp}^2)^2} \cdot
2\pi\delta(q_{1+}) .
\label{e42}
\end{equation}
It produces  a spinor factor $\gamma_{-}$ and a momentum factor which
allows us to represent the integral over longitudinal components
of the loop momentum $q_1$
in terms of $2I_2$ (see \Ref{e23}).

{\bf 1.} The spinor factor for the projectile quark for diagram 1
from Fig.~\ref{D1} is
\begin{equation}
\frac{i\gamma_{+}}{2} (i\hat{k}) \frac{i\gamma_{+}}{2} \cdot (-2i)
=-k_+ \gamma_{+} \ ,
\label{e44}
\end{equation}
where the factor $(-2i)$ is from the reggeon propagator.
The momentum factor for diagram 1 is
$$
-2I_2(k-p) L(p,q_1)
=
i\frac{L(p,q_1)}{4k_+}
$$
and for the diagram with exchanged reggeons is
$$
-2I_2(\lambda +p-k) L(p,q_3)
=
i\frac{L(p,q_3)}{4k_+} \ .
$$
The common color factor is
$$
\frac{1}{2N_c} f^{c b_1 a} T^{b_1} T^c
=-\frac{i}{4} T^a \ .
$$

{\bf 2.} The spinor factor for the projectile quark for diagram 2
from Fig.~\ref{D1} is the same as \Ref{e44}.
The momentum factor for diagram 2 is
$$
-2I_2(k) L(p,q_3)
=
i\frac{L(p,q_3)}{4k_+}
$$
and for the diagram with exchanged reggeons is
$$
-2I_2(\lambda -k) L(p,q_1)
=
i\frac{L(p,q_1)}{4k_+} \ .
$$
The common color factor is
$$
\frac{1}{2N_c} f^{c b_1 a} T^c T^{b_1}
=\frac{i}{4} T^a \ .
$$

{\bf 3.} The spinor factor for the projectile quark for diagram 3
from Fig.~\ref{D1} is
\begin{equation}
\frac{i\gamma_{+}}{2} \cdot (-2i)
=\gamma_{+} \ .
\label{e45}
\end{equation}
The momentum factor for diagram 3 is
$$
q_+ 2I_2(q) B(p,q_3,q_1)
=
-i\frac{B(p,q_3,q_1)}{4}
$$
and for the diagram with exchanged reggeons is
$$
q_+ 2I_2(\lambda -q) B(p,q_1,q_3)
=
-i\frac{B(p,q_1,q_3)}{4} \ .
$$
The common color factor is
$$
\frac{-i}{2N_c} f^{b b_1 d} f^{d b_1 a} T^b
=\frac{i}{2} T^a \ .
$$

From the relation $q_{1\perp}+q_{3\perp}=0$ the identities
\begin{equation}
B(p,q_1,q_3)=\frac{(p+q_1,\epsilon_\perp)}{(p+q_1)_\perp^2}
- \frac{(p+q_1 +q_3,\epsilon_\perp)}{(p+q_1 +q_3)_\perp^2}
=-L(p,q_1)
\label{e46}
\end{equation}
\begin{equation}
\mbox{ and } \quad
B(p,q_3,q_1)=-L(p,q_3)=-L(p,-q_1)
\label{e47}
\end{equation}
follow. As a result the total amplitude is found to be proportional to
\begin{equation}
-\frac{T^a}{8} \left( L(p,-q_1) + L(p,q_1) \right) ,
\label{e48}
\end{equation}
where the contribution of diagrams 1,2 and analogous ones with
exchanged reggeons cancels.
Collecting all factors we get for the diffractive amplitude
\begin{equation}
i g^5
(\gamma_{+}\otimes\gamma_{-})
\frac{T^a}{4}
\int\frac{d^2 q_{1\perp}}{(2\pi)^2} \frac{L(p,q_1)}{(q_{1\perp}^2)^2}\ .
\label{e49}
\end{equation}

The cut line of the projectile quark with momentum
$k'=k-(p+\lambda)$ contributes factor
\begin{equation}
2\pi \delta((k-p-\lambda)^2)
=
2\pi\delta(-k_+ \lambda_-)
=2\pi \delta(2k\lambda),
\label{e410}
\end{equation}
so the correspondent contribution to $F$ equals to just
the squared modulus of the amplitude.
The color factor is
\begin{equation}
\frac{1}{N_c} \tr(\frac{T^{a}}{4} \frac{T^{a}}{4})
=\frac{N_c^2-1}{32N_c} \ ,
\label{e4c}
\end{equation}
the spinor structure for the projectile quark is
\begin{equation}
\frac{1}{2}\sum_{\sigma} \bar{u}_{\sigma}({\bf k}) \gamma_{+}
\cdot \hat{k'} \cdot \gamma_{+} u_{\sigma}({\bf k})
\approx 4k_{+}^2 =4s .
\label{e411}
\end{equation}
One have to consider the two targets in the amplitude
and its complex conjugated as different, so
for both target quarks we have
\begin{equation}
\frac{1}{2}\sum_{\sigma} \bar{u}_{\sigma}({\bf l})
\gamma_{-} u_{\sigma}({\bf l})
=\frac{1}{2}\tr(\hat{l}\gamma_{-})
= 2l_{-} = 2\sqrt{s} \ .
\label{e412}
\end{equation}
If we denote $q_2$ the integration variable in the conjugated amplitude,
we find the total diffractive contribution as
\begin{equation}
F_9=\frac{1}{2} s^2 g^{10}\frac{N_c^2-1}{N_c}
L(p,q_1)L(p,q_2) ,
\label{e4f9}
\end{equation}
where the transverse integration \Ref{e27} is implied.

\section{Inclusive cross-sections}
\subsection{The total high-energy part $F$}

Collecting all results one can observe essential cancellations
between different contributions. Since
\begin{equation}
F_2+F_5=0, \quad
F_3+F_9=0, \quad
F_4+F_6=F_8 \ ,
\label{e51}
\end{equation}
the total coefficient is
\begin{equation}
F= F_1 + F_7 + 2F_8
=
s^2 g^{10}\frac{N_c^2-1}{N_c}
\left[
B^2(p,q_2,q_1)
-L^2(p,q_2)
+2L(p,q_2)B(p,q_2,q_1)
\right] \ .
\label{e5f1}
\end{equation}
Note that the obvious symetry of the transverse integral \Ref{e27}
with respect to the exchange $q_{1\perp}\lra q_{2\perp}$ is used
when needed.
Using the definition of vertices
we get the relation
\begin{equation}
B(p,q_2,q_1)+L(p,q_2)=L(p,q_{12}) \ ,
\quad \mbox{ where } q_{12}=q_1 + q_2 \ ,
\label{e52}
\end{equation}
which allows us to simplify the expression
\begin{equation}
F= s^2 g^{10}\frac{N_c^2-1}{N_c}
\left[
L^2(p,q_{12}) - 2 L^2(p,q_2)
\right] \ .
\label{e5f2}
\end{equation}

To calculate the cross-section one have to sum over polarizations
of the emitted gluon. Note that the light-cone gauge $\epsilon l=0$
(equivalent to $\epsilon_{+}=0$) was used in \cite{BV}-\cite{BSV}
and is implied in \Ref{evl} and \Ref{evb}.
For this gauge the spin sum
\begin{equation}
\sum_{\lambda}
\epsilon^{\lambda}_{\mu}(p) \epsilon^{\lambda}_{\nu}(p)
= -g_{\mu\nu} + \frac{p_{\mu}l_{\nu}+p_{\nu}l_{\mu}}{(pl)} \ ,
\label{e53}
\end{equation}
turns to $-g_{\mu\nu}$ for purely transverse components.
Then the contribution of the term $L^2(p,q)$ is
\begin{equation}
-\frac{1}{p^2_{\perp}} - \frac{1}{(p+q)^2_{\perp}}
+2 \frac{(p,p+q)_{\perp}}{p^2_{\perp} (p+q)^2_{\perp}}
=
-\frac{q^2_{\perp}}{p^2_{\perp} (p+q)^2_{\perp}} \ ,
\label{e54}
\end{equation}
so from \Ref{e5f2} we get
\begin{equation}
F= -s^2 g^{10}\frac{N_c^2-1}{N_c}
\left[
\frac{q^2_{12\perp}}{p^2_{\perp} (p+q_{12})^2_{\perp}}
-2\frac{q^2_{2\perp}}{p^2_{\perp} (p+q_2)^2_{\perp}}
\right] \ .
\label{e5f3}
\end{equation}
The second term in the brackets depends on only $q_2$.
Such contributions involve the pomeron at zero distance
between the reggeized gluons and vanish.
This brings us to the final result
\begin{equation}
F= -s^2 g^{10} N_c
\frac{q^2_{12\perp}}{p^2_{\perp} (p-q_{12})^2_{\perp}}\ ,
\label{e5ff}
\end{equation}
where we change the sign of integration variables $q_1,q_2$
and pass to the large $N_c$ limit to compare the expression
with one of the dipole picture.

Note that the same expression was obtained in ~\cite{braun3} within
the purely transverse picture assuming the validity of the AGK rules
for the relation of the diffractive, single cut and double cut
contributions as 1:-1:2. Our present derivation thus demonstrates
the validity of these rules.

\subsection{Impulse approximation}

We define the inclusive cross-section as
\beq
J(p)=\frac{(2\pi)^2d\sigma}{d^2pdy}
\eeq
leaving the additional factor $1/4\pi$ for the diagram.

We start from the impulse approximation, in which
\beq
J^{(1)}_A=AJ
\eeq
and $J$ is the cross-section on the nucleon ($J^{(1)}_A$ corresponds to
the cut diagram shown in Appendix 1 in Fig. \ref{fig8ap}).
As a diagram it carries momentum factor $[4(kl)]^2$
and colour factor $N_c^3$.
The cut couplings to the projectile and target give two $\delta$-functions
\[(2\pi)^2\delta(k_+q_-)\delta(l_-q_+)\]
which allow to do the longitudinal integrations and produce
factor $1/4(kl)$. Each emission vertex gives $2L(p,q)$ and we get
\[{\rm Im}{\cal A}=2(kl)N_c^3\frac{g^6}{4\pi}\int \frac{d^2q}{(2\pi)^2}
\frac{L^2(p,q)}{q^4}\]
(additional $1/2$ comes because we take the imaginary part and not twice it,
the latter corresponding to unitarity).
Including factor $2g^2N_c/4\pi$ into the BFKL interaction and dividing
by $s$ we find
\beq
J^{(1)}_A
%=-AN_c^2g^4\frac{d^2q}{(2\pi)^2q^4} V(-q,q|-q-p,q+p)
=
N_c^2g^4\frac{\alpha_sN_c}{p^2}
\int\frac{d^2q}{(2\pi)^2}\frac{1}{q^4}\frac{q^2}{(p-q)^2} \ ,
\eeq
where $p$ and $q$ denote transverse momenta in the Euclidean metric
that is with $p^2\geq 0$.

To interpret this expression we note that
\[\frac{g^2N_c}{2q^4}\equiv P^{(0)}_y(q)\]
give the lowest approximation for the pomeron $P_y(q)$ attached to the
target.
A similar lowest order pomeron attached to the projectile is
\[P^{(0)}_{Y-y}(p-q)=\frac{g^2N_c}{2(p-q)^4}\]
where $Y$ is the overall rapidity $\ln s$.
So we can rewrite
\beq
J^{(1)}_A=A\frac{4\alpha_sN_c}{p^2}
\int\frac{d^2q}{(2\pi)^2}(p-q)^2P_{Y-y}^{(0)}(p-q)q^2P^{(0)}_y(q)
\eeq
or in the coordinate space
\beq
J^{(1)}_A=A\frac{4\alpha_sN_c}{p^2}
\int d^2r e^{ipr}
\Big(\Delta P_{Y-y}^{(0)}(r)\Big)\Big(\Delta P^{(0)}_y(r)\Big) .
\eeq
In higher orders we obtain
\beq
J^{(1)}_A=A\frac{4\alpha_sN_c}{p^2}
\int d^2r e^{ipr}
\Big(\Delta P_{Y-y}(r)\Big)\Big(\Delta P_y(r)\Big)
\eeq
which is the standard expression for the gluon
emission from the BFKL ladder.

\subsection{Double scattering}

Now we have
\beq
J^{(2)}=\frac{1}{s^2}F\frac{1}{2}A(A-1)\int d^2b T_A^2(b) .
\eeq
Restoring the coefficients and transverse integration we have
in the Euclidean metric
\beq
F=
4(kl)^2 N_c^3g^8 \frac{\alpha_sN_c}{p^2}\int
\frac{d^2q_1d^2q_2}{(2\pi)^4}\frac{1}{q_1^4q_2^4}
\frac{q_{12}^2}{(p-q_{12})^2} \ .
\eeq

We interprete
\[
\frac{g^2N_c}{2q_1^4}=P^{(0)}_y(q_1),\ \ \frac{g^2N_c}{2q_2^4}=P^{(0)}_y(q_2)
\]
and as before introduce
\[
P^{(0)}_{Y-y}(p-q_{12})=\frac{g^2N_c}{2(p-q_{12})^4} \ .
\]
Then we find
\beq
F=2s^2g^2\frac{4\alpha_sN_c}{p^2}
\int
\frac{d^2q_1d^2q^2}{(2\pi)^4}
q_{12}^2(q_{12}-p)^2P^{(0)}_{Y-y}(p-q_{12})P^{(0)}_y(q_1)P^{(0)}_y(q_2) .
\eeq

We transform this expression to the coordinate space.
Passing to integration variables $q_{12}$ and $q_1$ with
$q_2=q_1-q_{12}$ we have the integral over $q_1$
\[
\int\frac{d^2q_1}{(2\pi)^2}\int d_2r_1d^2r_2
e^{iq_1r_1+i(q_1-q_{12})r_2}P_y^{(0)}(r_1)P^{(0)}_y(r_2)=
\int d^2re^{iq_{12}r}P_y^{(0)}(r)P^{(0)}_y(-r) .
\]
Multiplying this by $q_{12}^2$ we get
\[
-\int d^2re^{iq_{12}r}\Delta\Big(P_y^{(0)}(r)P^{(0)}_y(-r)\Big) .
\]
As a result
\[F=2s^2g^2\frac{4\alpha_sN_c}{p^2}
\int
\frac{d^2q_1d^2}{(2\pi)^2}d^2rd^2r'e^{i(p-q_{12})r'}
\Delta' P^{(0)}_{Y-y}(r')
e^{iq_{12}r}\Delta\Big(P_y^{(0)}(r)P^{(0)}_y(-r)\Big)\]\[=
s^2g^2\frac{4\alpha_sN_c}{p^2}
\int d^2r\Big(\Delta P^{(0)}_{Y-y}(r)\Big)
\Delta\Big(P_y^{(0)}(r)P^{(0)}_y(-r)\Big) . \]
Dividing by $s^2$ we get the inclusive cross section
\[
J^{(2)}_A=2g^2\frac{4\alpha_sN_c}{p^2}\rho^{(2)}
\int d^2r\Big(\Delta P^{(0)}_{Y-y}(r)\Big)
\Delta\Big(P_y^{(0)}(r)P^{(0)}_y(-r)\Big) \]
\beq
=
g^2\frac{4\alpha_sN_c}{p^2}A(A-1)
\int d^2r\Big(\Delta P^{(0)}_{Y-y}(r)\Big)
\Delta\Big(P_y^{(0)}(r)P^{(0)}_y(-r)\Big)
\int d^2b T_A^2(b) .
\eeq

\subsection{Comparison with the dipole picture}

We have obtained in the lowest approximation the total contribution
\beq
J_A=J^{(1)}_A+J^{(2)}_A=\frac{4\alpha_sN_c}{p^2}
\int d^2b
\int d^2r e^{ipr}
\Big(\Delta P_{Y-y}^{(0)}(r)\Big)\Delta \Big[P^{(0)}_y(r)T_A(b)+g^2
\Big(P^{(0)}_y(r)T_A(b)\Big)^2\Big] .
\eeq

In the dipole picture the inclusive cross-section has the form (~\cite{KT}
with some trivial redefinitions)
\beq
J_A=\frac{2\alpha_sN_c}{p^2}
\int d^2b
\int d^2r e^{ipr}
\Big(\Delta Q_{Y-y}(r)\Delta \Big[2{\cal N}_y(r,b)-{\cal N}^2_y(r,b)\Big]
\label{dipole}
\eeq
where
\[Q_y=-\frac{1}{g^2}P^{(0)}T_A(b)\]
and
 ${\cal N}(r,b)$ is the solution of the Balitsky-Kovchegov equation
\cite{bal,kov} with the initial condition
\beq
{\cal N}^{(0)}(r,b)=1-e^{g^2P^{(0)}(r)T_A(b)}
\eeq
so that in the two lowest orders in $g^2$ (evolution starts from
terms of the order $\alpha_s{\cal N}^2\sim g^6$)
\beq
{\cal N}(r,b)=-g^2P^{(0)}(r)T_A(b)-g^4\Big(P^{(0)}(r,b)T_A(b)\Big)^2
\eeq
and
\beq
2{\cal N}^{(0)}-{\cal N}^2=
-2g^2\Big[P^{(0)}(r)T_A(b)+g^2\Big(P^{(0)}(r,b)T_A(b)\Big)^2\Big] .
\eeq
One observes that in the lowest approximation
the dipole cross-section (\ref{dipole}) coincides with
the one found in our picture.

\section{Conclusions}

The main result of our paper is  demonstration of the
validity of the AGK rules for the inclusive gluon production,
namely that the diffractive, single and double cut contribution are
related as $1:-1:2$. Taking this into account we have found that
in the lowest approximation the inclusive cross-section found in the
reggeized gluon technique is identical with the one
obtained in the dipole picture. The decisive instrument in this
derivation has been the effective QCD action in the Regge kinematics
~\cite{lipatov}.
As stated in the Introduction this is sufficient to establish
equivalence of the dipole inclusive cross-section with the one determined
within the BFKL-Bartels picture in all orders provided the inclusive
cross-section is derived from the intermediate states  present
in the forward elastic scattering amplitude in Fig. \ref{npap3}.

We have to note that precisely this derivation has been questioned in
~\cite{BSVC}. Its authors argue that since the structure in Fig.~\ref{npap3}
is obtained from Feynman diagrams with the use of the so-called bootstrap
relation, the inclusive cross-section found directly from Feynman diagrams
may be different from the one found from
Fig. \ref{npap3} because probably the bootstrap  does not work for
inclusive diagrams.
In other terms the inclusive cross-section may contain additional terms
which do not change the total cross-section (integrate to zero over the
gluon momentum). According to ~\cite{BSVC} these additional terms are to
appear in the next-to-leading order in the inclusive cross-section.
However direct calculation of the next-to-leading order
performed in ~\cite{braun2} demonstrated that such terms
do not appear and the cross-section in the interacting reggeized gluon
framework continues to coincide with the dipole picture. This makes us
believe that the procedure to seek the emitted intermediate gluon directly in
Fig.~\ref{npap3} is correct, so that the equivalence of both approaches
is true.

\section{Acknowledgements}

This work has been supported by the RFFI grant 12-02-00356-a
and the SPbSU grants 11.059.2010 and 11.38.31.2011.

\section{Appendix 1. Heavy nucleus in the Glauber approach}

\subsection{Relativistic $AN^A$ vertex and the nucleus wave function}

\begin{figure}
\begin{center}
\includegraphics[scale=0.85]{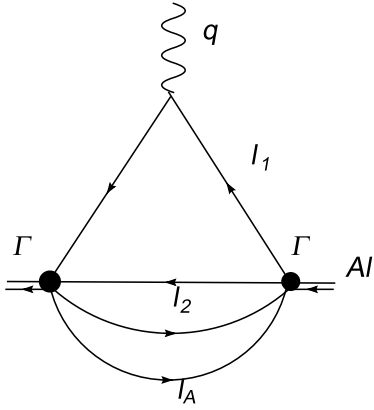}
\end{center}
\caption{Electromagnetic form factor of the nucleus.}
\label{fig7ap}
\end{figure}

To formulate the Glauber approach to collisions with a heavy nucleus
in the diagrammatic technique we first have to relate the relativistic
$AN^A$ vertex $\Gamma$ with the nucleus wave function. To this end we
study the baryonic form-factor of the nucleus illustrated in
Fig.~\ref{fig7ap} where vertices $\Gamma$ are shown with blobs.
In the lab. system and at zero momentum transfer it is equal
to $2AM$ where $M=A(m-\epsilon)$ is the nucleus mass. So we get the
normalization condition
\beq
A\int \prod_{j=2}^A\frac{d^4l_j}{((2\pi)^4 i(m^2-l_j^2-i0)}
\frac {2l_{10}\Gamma^2({\bf l}_j)}
{(m^2-l_1^2-i0)^2}=2AM\ .
\label{normnuc}
\eeq
Here $\sum_1^Al_j=Al$ and $A^2l^2=M^2$. For simplicity we neglect  spins and
consider all particles as scalar.
In the nucleus rest system we have $l_0=m-\epsilon$,
and ${\bf l}=0$, so that putting $l_j=
l+\lambda_j$, $j=2,...A$ we find
\[m^2-l_j^2=2m\epsilon-2m\lambda_{j0}-\lambda_{j\perp}^2,\ \
m^2-l_1^2=2m\epsilon+2m\sum_2^A\lambda_{0j}-\lambda_{1\perp}^2\ ,\]
where we used the orders of magnitude
$\lambda_0\sim\epsilon$, $|\lambda_\perp|\sim\sqrt{m\epsilon}$.
Integrations over $\lambda_{j0}$, $j=2,...A$ transform (\ref{normnuc}) into
\beq
\Big(\frac{1}{2m(2\pi)^3}\Big)^{A-1}\int \prod_{j=2}^Ad^3l_2
\frac {\Gamma^2({\bf l}_2,...{\bf l}_A)}
{(2Am\epsilon+\sum_1^A{\bf l}_j^2)^2}=A\ .
\label{normnuc1}
\eeq
Comparing with the standard normalization of the
nucleus wave function $\psi_A({\bf l}_j)$ we find the desired
relation
\beq
\psi_A({\bf l}_j)=\frac{1}{\sqrt{A}}\Big(\frac{1}{2m(2\pi)^3}\Big)^{(A-1)/2}
\frac{\Gamma({\bf l}_j)}{
2Am\epsilon+\sum_1^A{\bf l}_j}\ ,
\label{psinuc}
\eeq
which allows to relate the relativistic $AN^A$ vertex with the
nucleus wave function in the momentum space.

\begin{figure}
\begin{center}
\includegraphics[scale=0.85]{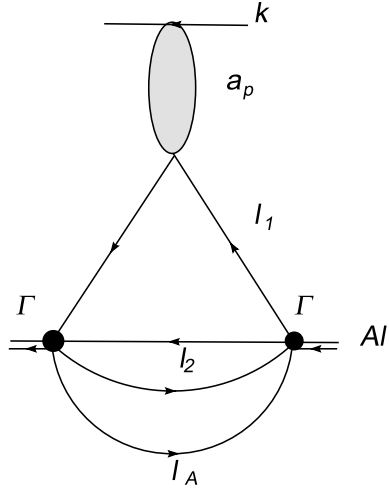}
\end{center}
\caption{Scattering on the nucleus in the impulse approximation.}
\label{fig8ap}
\end{figure}

To see how this works consider scattering on the nucleus in the
impulse approximation, Fig. \ref{fig8ap}.

The corresponding amplitude is given by
\beq
{\cal A}^{imp}=Aa_N\int \prod_{j=2}^A\frac{d^4l_j}{(2\pi)^4 i
(m^2-l_j^2-i0)}
\frac {\Gamma^2({\bf l}_2,...{\bf l}_A)}
{(m^2-l_1^2-i0)^2}\ ,
\eeq
where $a_N$ is the forward scattering amplitude off the nucleon.
Using (\ref{normnuc}) we find that the integral is equal to $A$ and
we get
$
{\cal A}^{imp}=A^2a_N.
$
But the relativistic flux on the nucleus is $A$ times that on the proton,
so that we get
$\sigma_A=A\sigma_N$, as expected.

\subsection{Double scattering on the nucleus}

\begin{figure}
\begin{center}
\includegraphics[scale=0.85]{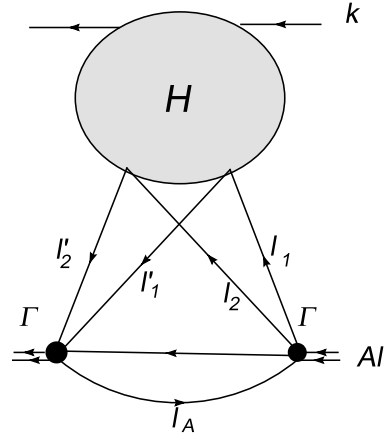}
\end{center}
\caption{Double scattering on the nucleus.}
\label{fig9ap}
\end{figure}

Now consider double scattering on the nucleus illustrated
in Fig.~\ref{fig9ap}.
Now the amplitude is given by
\[
{\cal A}=A(A-1)\int\prod_{j=3}^A\frac{d^4l_j}{(2\pi)^4 i(m^2-l_j^2-i0)}
\frac{d^4l_2}{(2\pi)^4 i}\frac{d^4l'_2}{(2\pi)^4 i}
H(l_{2z}-l'_{2z})\]\beq \times
\frac{\Gamma({\bf l}_2,{\bf l}_3,...{\bf l}_A)}{(m^2-l_1^2-i0)(m^2-l_2^2-i0)}
\frac{\Gamma({\bf l'}_2,{\bf l}_3,...{\bf l}_A)}
{(m^2-{l'_1}^2-i0)(m^2-{l'_2}^2-i0)}\ .
\label{doublenuc}
\eeq
Here $H$ is the high-energy part and it is taken into account that
it can only depend on the $z$-component of the transferred momentum,
since it is the only of the spatial components which enters multiplied
by the high projectile momentum. Factor $A(A-1)$ combines
factor 2 from  two different
ways in which $H$ can be coupled to the nucleus, corresponding to the
change $l'_1\lra l'_2$ and the choice of the active pair, which contributes
$(1/2)A(A-1)$.

Integrations over zero components of the nuclear momenta factorize
and taking into account (\ref{psinuc}) we get
\beq
{\cal A}=\frac{A^2(A-1}{2m(2\pi)^3}\int d^3l_2d^3l'_2\prod_{j=3}^A d^3l_j
H(l_{2z}-l'_{2z})
\psi_A({\bf l}_2,{\bf l}_3,...{\bf l}_A)
\psi_A({\bf l}'_2,{\bf l}_3,...{\bf l}_A)\ .
\label{doublenuc1}
\eeq

We rewrite
\[\int d^3l_2d^3l'_2\prod_{j=3}^A d^3l_j
H(l_{2z}-l'_{2z})
\psi_A({\bf l}_2,{\bf l}_3,...{\bf l}_A)
\psi_A({\bf l}'_2,{\bf l}_3,...{\bf l}_A)\]\[=
\int d^3l_1d^3l'_1 d^3l_2d^3l'_2\delta^3(l_1+l_2-l'_1-l'_2)
\prod_{j=3}^{A-1} d^3l_j
\psi_A(\bl_1,{\bf l}_2,{\bf l}_3,...{\bf l}_{A-1})
\psi_A({\bl'_1,\bf l}'_2,{\bf l}_3,...{\bf l}_{A-1})\]\[=
\int d^3l_1d^3l'_1 d^3l_2d^3l'_2\delta^3(l_1+l_2-l'_1-l'_2)
\rho(\bl_1,\bl_2|\bl'_1,\bl'_2)\ ,
\]
where now $\bl_A=\sum_1^{A-1}\bl_j$ and $(1/2)A(A-1)\rho_A$ is the density matrix
for the two active nucleons in the nucleus with $A$ nucleons.
So the amplitude becomes
\beq
{\cal A}=\frac{A}{m(2\pi)^3}
\int d^3l_1d^3l'_1 d^3l_2d^3l'_2\delta^3(l_1+l_2-l'_1-l'_2)
\rho_A(\bl_1,\bl_2|\bl'_1,\bl'_2)H(l_{2z}-l'_{2z})\ .
\eeq

To do the integrations over the transverse momenta we present
\[
\delta^2_\perp(l_1+l_2-l'_1-l'_2)=
\frac{1}{(2\pi)^2}\int d^2b e^{-i\bb (\bl_1+\bl_2-\bl'_1-\bl'_2)}
\]
and
\[
\rho_A(\bl_1,\bl_2|\bl'_1,\bl'-2)=\frac{1}{(2\pi)^4}
\int d^2b_1d^2b_2d^2b'_1d^2b'_2
\rho_A(\bl_{1\perp},l_{1z};\bl_{2\perp},l_{2z}|\bl'_{1\perp},l'_{1z}
\bl'_{2\perp},l'_{2z})\]\[
e^{-i\bb_1\bl_{1\perp}-i\bb_1\bl_{2\perp}+i\bb'_1\bl'_{1\perp} +i\bb'_2\bl'_{2\perp}}
\ . \]
Then after transversal integrations we get
\beq
{\cal A}=\frac{A}{2m\pi}
\int dl_{1z}dl'_{1z} dl_{2z}dl'_{2z}\delta(l_{1z}+l_{2z}-l'_{1z}-l'_{2z})
d^2b H(l_{2z}-l'_{2z})
\rho_A(\bb,l_{1z};\bb,l_{2z}|\bb,l'_{1z},\bb,l'_{2z})\ .
\eeq

Now we pass to spatial $z$ components and introduce the
$z$-component of the transferred momentum $\lambda_z=l_{2z}-l'_{2z}$.
Presenting
\[
\int dl_{1z}dl'_{1z} dl_{2z}dl'_{2z}\delta(l_{1z}+l_{2z}-l'_{1z}-l'_{2z})=
\int dl_{1z} dl_{2z}d\lambda_z
\]
we finally find

\beq
{\cal A}=\frac{A}{2\pi m}\int dz_1dz_2 d\lambda_z e^{-i\lambda_z(z_1-z_2)}
H(\lambda_z)
e^{-i\lambda_z z}\rho_A(\bb,z_1;\bb,z_2|\bb,z_1,\bb,z_2)\ .
\label{doublenuc5}
\eeq

The Glauber approximation follows if $H(\lambda_z)$ has a singularity
at $\lambda_z=0$. Typically
\beq
{\rm Im}\,H(\lambda_z)=2\pi\delta(2(k\lambda))F
=F\frac{\pi}{k_0}\delta(\lambda_z) .
\label{handf1}
\eeq
Here we use $\lambda_0<<|\lambda_z|$ and $k_0=k_z>0$.
In this case we get the Glauber approximation for the imaginary part of the
double scattering amplitude
\beq
{\rm Im}{\cal A}=\frac{A}{s}F\rho^{(2)}
\label{doublenuc6}
\eeq
where
\beq
\rho^{(2)}=\frac{1}{2}A(A-1)\int d^2b T^2_A(\bb) .
\eeq
Dividing by the flux $As$ we get the total cross-section
\beq
\sigma_A=\frac{1}{s^2}F\rho^{(2)} \ .
\eeq

\begin{figure}
\begin{center}
\includegraphics[scale=0.85]{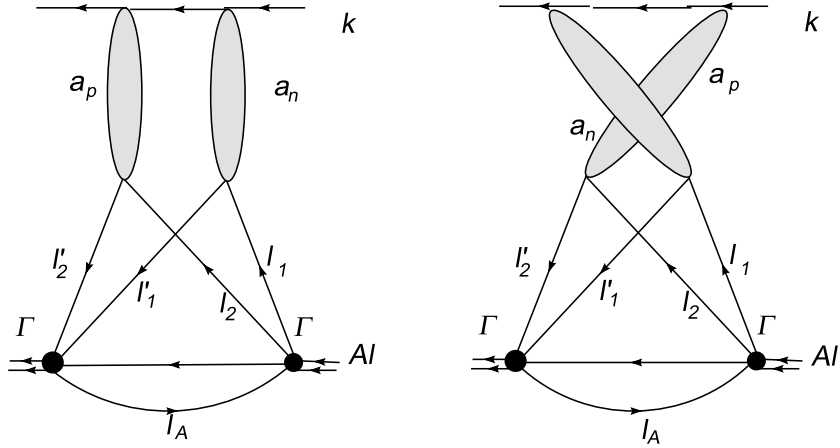}
\end{center}
\caption{Double elastic collision on the nucleus.}
\label{fig10ap}
\end{figure}

To see how this formula works consider again the simplest case of the
double scattering corresponding to double elastic collision
shown in Fig. \ref{fig10ap}.
In this case the  direct and crossed diagrams in Fig. \ref{fig10ap}
 give
\beq
H(\lambda_z)=2\pi i\delta(2k_0\lambda_z)a^2_N\ ,
\label{hsimplenuc}
\eeq
so that $F=ia^2_N$
and
\beq
{\cal A}=i\frac{A^2(A-1)}{2 s}a^2_N
\int d^2b T^2_A(\bb) .
\label{doublenuc7}
\eeq
The flux is $A$ times greater than off the nucleon. Dividing by $As$ we find
\beq
\sigma^{double}_A=\frac{1}{2}A(A-1)\sigma_N^2\int d^2b T^2_A(\bb)
\eeq
which is the standard expression for the double scattering in the Glauber
approximation.

The same derivation remains valid for the inclusive cross-section with the
obvious result
\beq
\frac{(2\pi)^3 d\sigma_A}{d^3p}=\frac{1}{s^2}
F\rho^{(2)}
\eeq
where of course $F$ refers to the diagrams with the external emitted
gluon of momentum ${\bf p}$.

\section{Appendix 2. Inclusive cross-section from the BFKL chain}

\begin{figure}
\begin{center}
\includegraphics[scale=1.0]{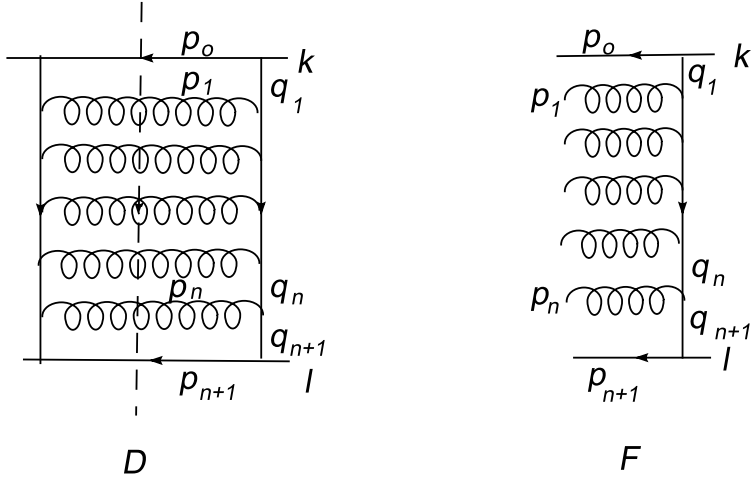}
\end{center}
\caption{Scattering amplitude D and production amplitude F with
$n$ real gluons.}
\label{npap1}
\end{figure}

Consider the imaginary part $D$ of the forward amplitude
for quark-quark scattering with
$n$ real gluons in the intermediate state  corresponding to the diagram
shown in Fig. \ref{npap1}, $D$. It is given by the square modulus of
the amplitude for the production of $n$ gluons $ F$ shown in Fig.
\ref{npap1},$F$.

Explicitly $2{\rm Im}\, D_n$ is given by the expression
\[2{\rm Im}\, D_n={g^2N_c}^{n+2}
\int\prod_{i=0}^{n+1}\frac{dp_{i+}}{4\pi p_{i+}}
(2\pi)^2\delta(\sum_{i=0}^{n+1}p_{i+}-k_+)
\delta(\sum_{i=0}^{n+1}p_{i+}-k_+)
\int\prod_{i=1}^{n+1}\frac{d^2q_{i\perp}}{(2\pi)^2}
|F_n|^2\ .\]
The production amplitude $F_n$ is given by
\[F_n=2s\frac{1}{\tilde{t}_{n+1}}\prod_{i=1}^n\frac{2 L(p_i,q_{i+1})}
{\tilde{t}_i}\ ,\]
where
\[\frac{1}{\tilde{t}_i}=\frac{s_{i,i-1}^{\omega(t_i)}}{t_i}\]
with $t_i=-|q_i^2|$ and $ L$ is the Lipatov vertex.

Apart from factor $1/(2\pi)^2$ associated with the
transverse integration, each intermediate gluon line carries factor
$g^2N_c/\pi$. Summing over polarizations with the
 factor $2g^2N_c/4\pi=2\alpha_s$ converts
$4 L^2(p_i,q_{i+1})$ into
 the standard BFKL interaction
\[-V(-q_{i+1},q_{i+1}|-q_{i+1}-p_1,q_{i+1}+p)\ .
\]
We are left with the additional factor $1/q_{n+1}^4$ for the last reggeon
and factors $s^{\omega(t_i)}$ in all reggeon propagators.
Note that all
factors $q_i^4$ except the last are included in $V$'s.
We denote
$|F_n|^2=4s^2T_n$ where $T_n$ is just the expression which corresponds
to the purely transverse picture.

The imaginary part of the amplitude $ D$ takes the form
\[2{\rm Im}\, D_n=\Big(\frac{g^2N_c}{4\pi}\Big)^2(2\pi)^24s^2
\int\prod_{i=0}^{n+1}\frac{dp_{i+}}{p_{i+}}
\delta(\sum_{i=0}^{n+1}p_{i+}-k_+)
\delta(\sum_{i=0}^{n+1}p_{i+}-k_+) T_D\equiv (g^2N_c s)^2 {\cal L}_nT_n
\ ,\]
where ${\cal L}_n$ is the longitudinal part
\[
{\cal L}_n=\int\prod_{i=0}^{n+1}\frac{dp_{i+}}{p_{i+}}
\delta(\sum_{i=0}^{n+1}p_{i+}-k_+)
\delta(\sum_{i=0}^{n+1}p_{i+}-k_+) . \]
In the Regge kinematics
\[ p_{0+}>>p_{1+}>>>....>>p_{n+}>>p_{n+1,+}\]
we can transform the $\delta$-functions into
\[\delta(p_{0+}-k_+)\delta(p_{n+1,-}-l_-)\]
and do the integrations over $p_{0+}$ and $p_{n+1,-}$.
We obtain
\beq
{\cal L}_n=\frac{2}{s}\int\prod_{i=1}^{n}\frac{dp_{i+}}{p_{i+}} \ .
\label{long}
\eeq

Next step is to transform the remaining
longitudinal phase volume to the variables introduced by Lipatov
$ s_{i,i-1}=2p_{i-}p_{i-1,+}$
to do the integrations by means of the two $\delta$-functions.

We arrange integrations as
\[
\int d\tau_n\equiv\int \frac{dp_{1,-}}{p_{1,-}}\int \frac{dp_{2,-}}{p_{2,-}}\int ...
\int \frac{dp_{n,-}}{p_{n,-}}\ .\]
Then we rewrite the integral as
\[
\int d\tau_n=\int \frac{d(2p_{0+}p_{1-})}{2p_{0+}p_{1+}}
\int \frac{d(2p_{1+}p_{2-})}{2p_{1+}p_{2-}}
...\int \frac{d(2p_{n-1,+}p_{n,-})}{2p_{n-1,+}p_{n,-}}=
\int \frac{\prod_{i=1}^nds_{i,i-1}}{\prod_{i=1}^ns_{i,i-1}} \ .\]
Now we use
\[\prod_{i=1}^{n+1}s_{i,i-1}=s\prod_{i=1}^n|p_{i\perp}|^2\]
and
\[
\int ds_{n+1,n}
\delta(\prod_{i=1}^{n+1}s_{i,i-1}-s\prod_{i=1}^n|p_{i\perp}|^2)
=\frac{1}{\prod_{i=1}^{n}s_{i,i-1}}
\]
to write the final expression for (\ref{long}) in the Lipatov form
\beq
{\cal L}_n=\frac{2}{s}\int\prod_{i=1}^{n+1}ds_{i,i-1}
\delta(\prod_{i=1}^{n+1}s_{i,i-1}-s\prod_{i=1}^n|p_{i\perp}|^2)\ .
\eeq

So we find
\beq
2{\rm Im}\,D_n=2s(g^2N_c )^2 T_n
\int\prod_{i=1}^{n+1}ds_{i,i-1}
\delta(\prod_{i=1}^{n+1}s_{i,i-1}-s\prod_{i=1}^n|p_{i\perp}|^2)\ .
\eeq

To pass to the inclusive cross-section we select some particular gluon
with momentum $p_m$, $1\leq m\leq n$, and split the integrands in both $T_n$
and ${\cal L}_n$ in two parts above and below the selected real gluon.
For the transverse part it is trivial
\[
T_n(q_1,..q_n)=\int \frac{d^2q_m}{(2\pi)^2}\frac{d^2q_{m+1}}
{(2\pi)^2}T_{m-1}(q_1,...q_{m-1})q_m^4
V(-q_{m+1}q_{m+1}|-q_{m},q_m)
T_{n-m}(q_{m+1},...q_n)\ ,
\]
where $q_{m+1}=q_m-p$.
Factor $q_{m}^4$ eliminates the redundant denominator $1/q_m^4$
which is contained in both $T_m$ and $V$.
To factorize the longitudinal part we use
\[
\delta(\prod_{i=1}^{n+1}{i,i-1}-s\prod_{i=1}^n|p_{i\perp}|^2)\]\[=
\int ds_1ds_2\delta(s_1s_2-s|p_m|^2)
\delta(\prod_{i=1}^{m}s_{i,i-1}-s_1\prod_{i=1}^{m-1}|p_{i\perp}|^2)
\delta(\prod_{i=m+1}^{n+1}s_{i,i-1}-s\prod_{i=m+1}^n|p_{i\perp}|^2) . \]

As a result we present $2{\rm Im}\,D_n$ in the factorized form
\[
2{\rm Im}\, D_n=2s(g^2N_c )^2\int
ds_1ds_2(s_1s_2-sp_m^2)
\frac{d^2q_m}{(2\pi)^2}\frac{d^2q_{m+1}}{(2\pi)^2}\]\[
\Big[q_m^4T_{m-1}(q_1,...q_{m-1})\int \prod_{i=1}^{m}ds_{i,i-1}
\delta(\prod_{i=1}^{m}s_{i,i-1}-s_1\prod_{i=1}^{m-1}|p_{i\perp}|^2)\Big]
\]\[
\Big[-V(-q_{m+1}q_{m+1}|-q_{m+1}-p_m,q_{m+1}+p)\Big]\]\[
\Big[T_{n-m}(q_{m+1},...q_n)\int \prod_{i=m+1}^{n+1}ds_{i,i-1}
\delta(\prod_{i=m+1}^{n+1}s_{i,i-1}-s\prod_{i=m+1}^n|p_{i\perp}|^2)\Big]
\ .\]
Fixing $p_m=p$ and summing over both $m$ and $n$ we finally find
\[
2{\rm Im}\, D_n=-2s(g^2N_c )^2\int
ds_1ds_2(s_1s_2-sp_m^2)
\frac{d^2p}{(2\pi)^2}\frac{d^2q}{(2\pi)^2}P_{s_1}(p+q)q^4P_{s_2}(q)
V(-q,q|-q-p,q+p)\ ,
\]
where
\beq
P_{s_1}(p+q)=g^2N_c\sum_m
\Big[T_{m-1}(q_1,...q_{m-1})\int \prod_{i=1}^{m}ds_{i,i-1}
\delta(\prod_{i=1}^{m}s_{i,i-1}-s_1\prod_{i=1}^{m-1}|p_{i\perp}|^2)\Big]
\eeq
with $q_m=p+q$ has the meaning of the pomeron with energy $\sqrt{s_1}$
attached to the projectile
and similarly $P_{s_2}(q)$ is the pomeron attached to the target.

The inclusive cross-section will be given by
\[
\frac{(2\pi)^2d\sigma}{d^2pdy}=-\int\frac{d^2q}{(2\pi)^2}
V(-q,q|-q-p,q+p)(p+q)^4P_{Y-y}(p+q)P_y(q)\ ,\]
or using the explicit expression for the BFKL interaction
\beq
\frac{(2\pi)^2d\sigma}{d^2pdy}=
\frac{4\alpha_sN_c}{p^2}\int
\frac{d^2q}{(2\pi)^2}
P_{Y-y}(p+q)q^2(p+q)^2P_y(q)
\eeq
which is the standard expression.

\end{document}